\documentclass[range]{ar2esinglesppreprnt}

\def\Htwo{{\hbox {H$_2$}}}

\def\Lco{{\hbox {$L_{\rm CO}$}}}

\def\Lfir{{\hbox {$L_{\rm FIR}$}}}
\def\Ico{{\hbox {$I_{\rm CO}$}}}

\def\Tmb{{\hbox {$T_{\rm mb}$}}}

\def \deg{^\circ}

\def \sqr#1#2{{\vcenter{\hrule height.#2pt
     \hbox{\vrule width.#2pt height#1pt \kern#1pt
         \vrule width.#2pt}
     \hrule height.#2pt}}}

\def\LprimeCO{{\hbox {$L^\prime_{\rm CO}$}}}
\def\Lprimeco{{\hbox {$L^\prime_{\rm CO}$}}}
\def\Lcop{{\hbox {$L^\prime_{\rm CO}$}}}
\def\Lhcnp{{\hbox {$L^\prime_{\rm HCN}$}}}
\def\lcop{{\hbox {$L^\prime_{\rm CO}$}}}
\def\Lco {{\hbox {$L_{\rm CO}$}}}
\def \Mdyn{{\hbox {$M_{\rm dyn}$}}}

\def \MH2{M(\Htwo)}
\def \MiH2{M_i(\Htwo)}
\def \nH2{n(\Htwo)}
\def \nbarH2{\bar{n}}
\def \niH2{n_i(\Htwo)}
\def \fH2{f(\Htwo)}
\def \H2{\Htwo}
\def\Htwo{{\hbox {H$_2$}}}
\def\htwo{{\hbox {H$_2$}}}

\def\,{\thinspace}
\def\Msun{M$_\odot$}
\def \msun {\hbox{M$_\odot$}}
\def \Lsun {\hbox{L$_\odot$}}
\def \lsun {\hbox{L$_\odot$}}
\def\Lfir{{\hbox {$L_{\rm FIR}$}}}
\def\lfir{{\hbox {$L_{\rm FIR}$}}}

\def \kms{{\hbox{km\,s$^{-1}$}}}
\def \Kkmspc{K\,\kms\,pc$^2$}
\def \kkmspc{K\,\kms\,pc$^2$}
\def \microns {$\mu$m}

\def \micron {$\mu$m}

\def \arcsec {^{\prime\prime}}

\def\Kkmspc{K\,\kms\,pc$^2$}

\def\Lcop{L'_{\rm CO}}

\def\vlap#1{\vbox to 0pt{\hbox{#1}\vss}}
\def\bull{\vrule height .9ex width .8ex depth -.1ex } 

\def\LprimeCO{{\hbox {$L^\prime_{\rm CO}$}}}
\def\Lprimeco{{\hbox {$L^\prime_{\rm CO}$}}}
\def\Lcop{{\hbox{$L^\prime_{\rm CO}$} }}
\def\Lco {  { \hbox { $L_{\rm CO}$} }  }
\def \Mdyn{{\hbox {$M_{\rm dyn}$}}}

\def \MH2{M(\Htwo)}
\def \MiH2{M_i(\Htwo)}
\def \nH2{n(\Htwo)}
\def \nbarH2{\bar{n}}
\def \niH2{n_i(\Htwo)}
\def \fH2{f(\Htwo)}
\def \H2{\Htwo}
\def\Htwo{{\hbox {H$_2$}}}
\def\htwo{{\hbox {H$_2$}}}

\def\,{\thinspace}
\def\Msun{M$_\odot$}
\def \msun {\hbox{M$_\odot$}}
\def \Lsun {\hbox{L$_\odot$}}
\def \lsun {\hbox{L$_\odot$}}
\def\Lfir{{\hbox {$L_{\rm FIR}$}}}
\def\lfir{{\hbox {$L_{\rm FIR}$}}}

\def \kms{{\hbox{km\,s$^{-1}$}}}
\def \Kkmspc{K\,\kms\,pc$^2$}
\def \kkmspc{K\,\kms\,pc$^2$}
\def \microns {$\mu$m}

\def \micron {$\mu$m}

\def \CI {\scshape Ci\rm}
\def \fCI {[\scshape Ci]\rm}
\def \TCI {[\scshape Ci]\rm}

\def \CIu {$^{3}P_{2}$}
\def \CIl {$^{3}P_{1}$}
\def \CIll {$^{3}P_{0}$}
\def \CIIu {$^{2}P_{3/2}$}
\def \CIIl {$^{2}P_{1/2}$}
\def \Lcpu {$L^\prime$\fCI(\CIu $\rightarrow$ \CIl)}
\def \Lcpl {$L^\prime$\fCI(\CIl $\rightarrow$ \CIll)}

\def \CII {\scshape Cii \rm}
\def \fCII {[\scshape Cii] \rm}

\usepackage{lscape}
\usepackage{multirow}
\usepackage{epsfig}
\begin{document}
{
\renewcommand{\arraystretch}{1.2}
\input epsf.tex    

\jname{..}
\jyear{2000}
\jvol{}
\ARinfo{1056-8700/97/0610-00}

\title{Molecular Gas at High Redshift}

\markboth{SOLOMON \bull \hspace{.1pt} VANDEN BOUT}{MOLECULAR GAS AT  
HIGH REDSHIFT}

\author{P. M. Solomon 
\affiliation{Dept. of Physics and Astronomy, State University of New York at Stony Brook, Stony Brook  
NY 11794 [psolomon@sbastk.ess.sunysb.edu]}
P. A. Vanden Bout 
\affiliation{National Radio Astronomy Observatory, 520 Edgemont Road,  
Charlottesville VA 22903 [pvandenb@nrao.edu]}}


\begin{abstract}
The Early Universe Molecular Emission Line Galaxies (EMGs) are a  
population of
galaxies with only 36 examples  that hold great promise for the study of
galaxy formation and evolution at high redshift.  The classification,  
luminosity of molecular line emission, molecular mass, far-infrared (FIR)
luminosity,  star formation efficiency,   morphology,
  and dynamical mass of the currently known sample are presented and
discussed.  The star formation rates derived from the FIR  
luminosity range
from about 300 to 5000 \Msun\ year$^{-1}$ and the molecular mass from  
$4\times10^{9}$
to $1\times10^{11}$ \Msun.
  At the lower end, these star formation rates, gas masses,  and  
diameters are
similar to those of local ultraluminous infrared galaxies, and
represent starbursts in centrally concentrated disks, sometimes, but not
always, associated with active galactic nuclei. The evidence for large ($> 5$ kpc) molecular disks
is limited.  Morphology  and several high angular resolution images  
suggest that
some   EMGs are mergers with a massive molecular interstellar
medium in both components.   A critical question is whether the EMGs, in
particular those at the higher end of the gas mass and luminosity  
distribution,
represent the formation of massive, giant elliptical  galaxies in the  
early
Universe.
  The sample size is expected to grow explosively in the era of the Atacama Large Millimeter Array (ALMA).    

\end{abstract}

\maketitle

\section{INTRODUCTION}

One of the many important advances in our knowledge of the distant,  
early Universe during the past
decade has come from  observations of spectral line emission from  
interstellar molecular gas, the
raw material from which stars form, in high-redshift ($z>2$) galaxies. For  
convenience, we call these
objects Early (Universe) Molecular (Line Emission) Galaxies, or EMGs.  
The molecular interstellar
medium (ISM) plays a critical role in the evolution of galaxies; these  
observations provide the first
evidence of the location and mass of molecular clouds during  the epoch 
  of galaxy formation. To
date, observations of rotational transitions of carbon monoxide (CO)  
have been reported for 36
sources with redshift $z > 1$, unequivocally demonstrating that  
molecular clouds, an extreme
Population I component, appeared early in the history of the Universe.  (For completeness, we have included three galaxies with CO detections at redshifts $1<z<2$ in this review.)  
  The jump  from detecting CO
in local
  ($z\leq 0.3$) galaxies to high-redshift  observations was made  
possible by the increased
sensitivity of millimeter-wave telescopes and arrays.  It was also  
facilitated by the large masses of
molecular gas associated with EMGs, a ``negative K-corrrection'' (see Section 2.1) for CO 
  emission, gravitational
lensing of many of the sources, and selection of sources with strong  
FIR emission,
which is often associated with star-forming molecular gas.

Almost all candidate galaxies successfully detected in 
high-redshift CO 
  emission were first identified
as strong FIR/submillimeter sources with FIR luminosity in excess of  
$10^{12}$ \Lsun. Given the
relatively narrow instantaneous bandwidth of millimeter-wave receivers  
and spectrometers, an
important selection criterion for CO emission line searches has been  
the availability of accurate
redshifts from optical line spectroscopy. This situation   
changes as instrumental
bandwidths increase.

Two main techniques have been applied to find most  EMGs.   
The first   employs large
optical surveys of bright high-$z$ quasars as a potential source list  
followed by observations of the
flux at 1.2 mm or 0.85 mm. At these wavelengths the continuum of an EMG 
  is dominated by thermal
dust emission rather than an extension of the nonthermal radio  
continuum.   CO emission  has now
been observed from 16 quasars, including the most distant known quasar  
at $z = 6.4$ (Walter et al. 2003). The second
technique identifies highly luminous infrared (IR) galaxies from blank  
field observations with
submillimeter-wave bolometers. Although not targeting individual cases of  
strong lensing, these
observations often take advantage of intermediate-redshift  cluster  
lensing. These techniques have
led to the discovery of extremely luminous dusty FIR galaxies at high  
redshift, similar to local
ultraluminous infrared galaxies (ULIRGs), but with a much higher space  
density. The search for CO in
these submillimeter galaxies (SMGs) illustrates the (historical)  
importance of having good redshifts.
Initial searches using Ly$\alpha$ redshifts were disappointing; later,  
the availability of H$\alpha$
redshifts led to a success rate of $>50\%$.   A total of 11  
SMGs have
been reported as having CO emission.  There are 73 SMGs with  
spectroscopic redshifts (Chapman et al. 2005), so a large
number of CO detections is possible in the surveys underway. A third  
detection strategy involved
searching IR-luminous radio galaxies for CO emission.  Seven such  
detections have been reported.
Finally, one Lyman Break galaxy (LBG) has been observed in CO emission, 
  a detection made possible
by strong magnification by a gravitational lens, and one extremely red  
object (ERO) has been
detected.

The discovery of high-redshift CO emission predates these surveys.   
IRAS F10214 was a source at
the detection limit of IRAS in the 60 and 100 $\mu$m bands, shown to be 
  of high FIR luminosity
when its redshift of $z=2.3$ was (serendipitously) measured  
(Rowan-Robinson et al. 1991).  The high
FIR luminosity motivated a successful search for the rotational J=3--2 line of CO with  
the NRAO 12m Telescope
(Brown and Vanden Bout 1991, 1992).   The (3--2) detection was soon confirmed
at the Institut Radioastronomie Millem\'etrique (IRAM) 30m Telescope, but with a much smaller flux  
(Solomon, Downes, \& Radford 1992a), and the CO(6--5) line
was also observed, indicating the presence of warm  molecular gas  
typically associated with
star formation.   Successful searches
for redshifted CO emission in several quasars soon followed: the  
Cloverleaf at $z=2.6$
(Barvainis et al. 1994), BR1202  at $z=4.7$ (Omont et al. 1996b), and BRI1335 at $z=4.4$  
(Guilloteau et al. 1997).

The CO observations of EMGs have the potential to answer several  
important questions about star
formation and galaxy evolution in the early Universe:  What is the mass 
  of molecular gas and how
does it compare with the dynamical mass? Are the EMGs centrally  
concentrated, as are most local
ULIRGs, or are they extended protogalaxies with substantially more  
molecular mass than that of ULIRGs?
What is the star formation lifetime? What is the
final evolutionary state of the EMGs?

\section{DEFINING THE EMGs}

\subsection{Luminosities: Basic Relations}

The calculation of high-redshift source properties from the observation
of molecular emission lines requires care with respect to the cosmology
assumed.  This is important when comparing published source properties,
as different cosmologies can lead to significantly different values for
properties such as luminosity, size, mass.   In this
review we have assumed a  cosmology with
$\Omega_{\rm m} = 0.3$, $\Omega_{\Lambda} = 0.7$, and $H_0 = 70$ km s$^{-1}$
Mpc$^{-1}$.

The CO line luminosity  can be expressed in several  ways. From
energy conservation, the monochromatic luminosity, observed flux  
density, and luminosity distance are related by
$\nu_{\rm rest} L(\nu_{\rm rest}) = 4 \pi D_{\rm L}^{2} \nu_{\rm obs}
S(\nu_{\rm obs})$, yielding
\begin{equation}
\Lco = 1.04 \times 10^{-3} \, S_{\rm CO}\, \Delta v\,
     \nu_{\rm rest} (1+z)^{-1}\, D_{\rm L}^{2},
\label{eq1}
\end{equation}
where the CO line luminosity, $\Lco$, is measured in \Lsun; the  
velocity integrated flux,
$S_{\rm CO}\, \Delta v$, in Jy\,\kms; the rest frequency, $\nu_{\rm  
rest} =
\nu_{\rm obs} (1+z)$, in GHz; and the luminosity distance,
$D_{\rm L}$, in Mpc.\footnote{The rough dependence of the luminosity 
distance on  redshift can be seen from the following: $D_{\rm L} = D_{\rm A} 
(1+z)^{2}$, where  $D_{\rm A}$ is the angular size distance.  For the 
cosmology assumed in this  review, $D_{\rm A}$ rapidly increases with 
redshift, reaching a peak value  at  $z\approx1.6$, and then declines 
roughly as $(1+z)^{-1}$ for larger $z$.  So for redshifts larger than 
$z\sim2$, $D_{\rm L}$ grows  roughly as $(1+z)$.  A calculator for computing  
luminosity and angular size distances in any cosmology can be found at  
http://www.astro.ucla.edu/$\sim$wright/CosmoCalc.html.}

The CO line luminosity is often expressed (Solomon et al. 1997) in units of  
\Kkmspc\ as the product of the
velocity integrated source brightness temperature, $T_{\rm b} \, \Delta 
  v$,
and the source area, $\Omega_{\rm s}  D_{\rm A}^{2}$, where $\Omega_{\rm s}$
  is the solid angle subtended by the source.
The observed integrated line
intensity, $\Ico = \int \Tmb\,dv$, measures the beam diluted brightness
temperature, which  decreases with redshift,
  $T_{\rm b} \, \Delta v\, \Omega_{\rm s}  = \Ico \Omega_{\rm s \star b}
(1+z)$, where $\Omega_{\rm s \star b}$ is the solid angle of the source
convolved with the telescope beam.  Then the line luminosity
$\Lcop = T_{\rm b} \, \Delta v \, \Omega_{\rm s} D_{\rm A}^{2}
= \Omega_{\rm s \star b} D_{\rm L}^{2} \Ico  (1+z)^{-3}$, or
\begin{equation}
  \Lcop = 23.5 \, \Omega_{\rm s \star b}\, D_{\rm L}^{2}\, \Ico\, (1+z)^{-3}
\label{eq2}
\end{equation}
where $\Lcop$ is measured in \Kkmspc, $\Omega_{\rm s \star b}$ in  
arcsec$^2$,
$D_{\rm L}$ in Mpc, and $\Ico$ in K\,\kms.  If
the source is much smaller than the beam, then  $\Omega_{\rm s \star b}  
\approx
\Omega_{\rm b}$.

The line luminosity, $\Lcop$, can also be expressed for a source of any 
  size
in terms of the total line flux, $\Lcop = (c^{2}/2k) S_{\rm CO}\,  
\Delta v \,
  \nu_{\rm obs}^{-2} \, D_{\rm L}^{2}\, (1+z)^{-3}$, or
\begin{equation}
\Lcop = 3.25 \times 10^{7} \, S_{\rm CO}\, \Delta v \,
     \nu_{\rm obs}^{-2}\, D_{\rm L}^{2} \, (1+z)^{-3}.
\label{eq3}
\end{equation}

Because  $\Lcop$ is proportional to brightness temperature, the $  
\Lcop$ ratio for two lines
in the same source is equal to the ratio of their intrinsic  brightness 
  temperatures averaged
over the source. These ratios provide important constraints on physical 
  conditions in the gas.
For thermalized optically thick CO emission the intrinsic brightness  
temperature and line
luminosity are independent of $J$ and of rest frequency.  For example,  
$\lcop
(J=3-2) = \lcop (J=1-0)$.

By observing CO emission from higher $J$ transitions for 
high- redshift  
galaxies researchers can
maintain the same approximate observed frequency as redshift increases. 
   Equations 2 and 3
show that for fixed line luminosity (\lcop) and a   fixed observed  
frequency (or a fixed beam
size), the observed integrated line intensity and the integrated flux   
do  not scale as the
inverse square of luminosity distance ($D_{\rm L}^{-2}$), but rather as  
$(1+z)^{3} D_{\rm L}^{-2}$ . This
substantial negative K-correction
(Solomon, Downes \& Radford 1992a,b)
  is one of the reasons  the relatively clear 3-mm atmospheric window,  
with instruments
developed for observation of CO(1--0) in the local Universe, has been  
the most important
wavelength band  for observations of  CO from EMGs at
$z \geq 2$.

A significant fraction of the EMGs are gravitationally imaged by an  
intervening galaxy.  The
luminosities $L$ and
$L'$ calculated without correction for the magnification by the  
gravitational lens are,
therefore, only apparent luminosities.  If a model of the gravitational 
  lens is available, the
intrinsic luminosities can be calculated from
$L_{\rm int}=L_{\rm app}/\mu$ and $L'_{\rm int}=L'_{\rm app}/\mu$, where $\mu$ is the  
area magnification factor
of the gravitational lens.  Wiklind
\& Alloin (2002) have reviewed gravitational lensing of EMGs.

\subsection{From CO Luminosity to Molecular Mass}

Observation of emission from CO rotational transitions is the dominant  
means of tracing
interstellar molecular clouds, which consist almost entirely of  
molecular hydrogen, \Htwo.
Molecular hydrogen rather than atomic hydrogen is the principal  
component of all interstellar
clouds with density $n > 100$ cm$^{-3}$ owing to a balance between  
formation on dust and
self-shielding of \Htwo\ from photodissociation (Solomon \& Wickramasinghe 1969) by the 
  interstellar radiation
field. This transition from atomic to molecular hydrogen at a moderate  
interstellar density
means that all dense clouds are molecular. Molecular clouds are  the  
raw material for star
formation and a critical component in the evolution of galaxies. The  
 first generation of stars
must have formed, in the absence of heavy elements, from HI with only  
trace amounts of
\Htwo\ available to provide essential cooling. However, the huge  
IR luminosity seen in
ULIRGs and EMGs is clearly emitted by interstellar dust, and we can  
expect all dense, dusty clouds
to be molecular.
\Htwo\  has  strongly forbidden rotational transitions, and the \Htwo\ vibration-rotation lines  
require high temperature to be
produced, for example, by UV excitation or shocks. In the absence of  
these special circumstances,
the \Htwo\ is invisible.

CO emission is the best tracer of molecular hydrogen for two reasons.  It is a very
stable molecule and the most abundant molecule after \Htwo. Second, a weak  dipole
moment ($\mu_{e}=0.11$ Debye) means that CO rotational levels are excited and 
thermalized by collisions with \Htwo\ at relatively low molecular hydrogen
densities.  Strong CO emission  from  interstellar gas dominated by
\Htwo\ is ubiquitous.  The critical density necessary to produce substantial excitation
of a rotational  transition is given approximately by  n(\Htwo ) $\geq  $ A/C where A
is the Einstein coefficient for spontaneous decay and C is the collisional rate
coefficient.  The A coefficient scales as $\mu^2\nu^3$ where $\mu$ is the dipole
moment and $\nu(J, J-1) = 2BJ$ for a simple rotational ladder,  is the frequency of the
transition . In practice the  critical density is lowered  by line trapping for CO
emission and for emission from other optically thick tracers  such as HCN and CS.
The full multi level excitation problem must be solved usually using the LVG (large
velocity gradient) approximation (Scoville and Solomon 1974; Goldreich and Kwan
1974).  The effective density for strong CO emission ranges from   n(\Htwo )
$\approx$  300 cm$^{-3}$ for  J = (1-0) to  
$\approx$  3000 cm$^{-3}$ for J = (4-3) or (5-4). Of course the higher J transitions
also require a minimum kinetic temperature for collisional excitation. 

 For high-$z$ galaxies there is another obvious  requirement for strong CO emission.
The large
quantities of dust and molecular gas observed in EMGs clearly indicate  
not only ongoing star
formation but also substantial enrichment by previous star formation. Researchers have known for some
time that many quasar emission line regions show substantial  
metallicity; EMGs have not only a
high metallicity, but also  a  huge mass of enriched interstellar  
matter much larger and more
extensive than that of a quasar emission line region.


  The H$_2$ mass-to-CO luminosity relation can be expressed as
\begin{equation}
M({\rm H}_2) = \alpha L^\prime_{\rm CO}\ \ ,
\label{eq4}
\end{equation}
  where $M({\rm H}_2)$ is defined to include the mass of He, so that  
$M({\rm H}_2) = M_{\rm gas}$,
the total gas mass, for molecular clouds.  For the Galaxy, three  
independent analyses yield the
same linear relation between the gas mass and the CO line luminosity:    
({\it a}) correlation of optical/IR
extinction with $^{13}$CO in nearby dark clouds (Dickman 1978);   ({\it b}) correlation of the flux of $\gamma$  
rays, produced by cosmic ray
interactions with protons, with the CO line flux for the Galactic  
molecular ring (Bloemen et al. 1986, Strong et al. 1988); and ({\it c}) the observed relations between virial  mass  
and CO line luminosity for
Galactic  giant molecular clouds (GMCs) (Solomon et al. 1987), corrected for 
  a solar circle radius of
8.5 kpc. All these methods  indicate that in our Galaxy,
$\alpha\equiv  M_{\rm gas}$/\Lcop = 4.6 \Msun  
(K\,km\,s$^{-1}$\,pc$^2)^{-1}$
(Solomon \& Barrett 1991). (Some authors use X rather than $\alpha$ as 
  a symbol for this
conversion factor, even though X by convention relates  \Htwo\ column  
density and
line-integrated CO intensity.)

For a single cloud or
an ensemble of nonoverlapping clouds, the gas mass determined from the 
  virial
theorem,
$M_{\rm gas}$, and the CO line luminosity, $\Lcop$, are
  related by
$$ \displaystyle
M_{\rm gas}\,/\,\Lcop = \alpha =
\left({{4\,m'}\times 1.36
\over {3\,\pi\,G}}
\right)^{1/2}
{n^{1/2}\over T_{\rm b}} \ \
= \ 2.6\
{n^{1/2}\over T_{\rm b}}
\eqno(5)
$$
where $m'$ is the mass of an H$_2$ molecule multiplied by 1.36 to account  
for He,
$n$(cm$^{-3}$) is the average H$_2$ number density in the clouds, and  
$T_{\rm b}$(K) is the
intrinsic (rest-frame) brightness temperature of the CO line.
$M_{\rm gas}$ is in \Msun\ and \Lcop\ is in \kkmspc\ (Dickman, Snell \& Schloerb 1987; Solomon et al. 1987). This is
the physical basis for deriving gas mass from CO luminosity.   The  
existence of gravitationally
bound clouds is confirmed by the agreement between $\alpha$ determined  
from application of
the virial theorem, using measured velocity dispersions and sizes for the  
Milky Way clouds, and
$\alpha$ determined from the totally  independent methods ({\it a}) and  
({\it b}) discussed above.

Use of the Milky Way value for the molecular gas mass to CO luminosity
  ratio, $\alpha$ = 4.6\,\Msun (\Kkmspc)$^{-1}$, overestimates the gas mass in  
ULIRGs  and probably in
EMGs. After high-resolution maps were produced for a few ULIRGs  
(Scoville, et al. 1991) it became
apparent that the molecular gas mass calculated using the Milky Way  
value for $\alpha$ was
comparable to and in some cases greater than the dynamical mass of the  
CO-emitting region.
This contradiction led to a new model (Downes, Solomon \& Radford 1993; Solomon et al. 1997) for CO 
  emission in ULIRGs.
Unlike Galactic clouds or gas distributed in the disks of galaxies,  
most of the CO emission in the
centers of ULIRGs may not come from many individual virialized clouds, but from a  
filled intercloud medium,  so
the linewidth is determined by the total dynamical mass in the region  
(gas and stars), that is,
$\Delta V^{2} = G\, \Mdyn /R $.  The CO luminosity
   depends on the  dynamical mass as well  as the gas mass. The CO line  
emission may trace a
medium bound by the total potential of the galactic center, containing  
a mass \Mdyn\ consisting
of stars, dense clumps, and an interclump medium; the interclump medium containing the CO  
emitting gas with mass
$M_{\rm gas}$.

Defining $f \equiv M_{\rm gas}/M_{\rm dyn} $,
the usual CO to \htwo\  mass relation becomes (Downes, Solomon \& Radford 1993)
$$M_{\rm dyn}/\LprimeCO  \ = \ f^{-1/2} \,\alpha\,
         \ = \ f^{-1/2} \,2.6\,(\nbarH2)^{1/2}\,T_{\rm b}^{-1}
\ \ ,$$
$$ M_{gas}/\LprimeCO  \ = \ f^{1/2} \,\alpha\,
         \ = \ f^{1/2} \,2.6\,(\nbarH2)^{1/2}\,T_{\rm b}^{-1}
\ \ ,$$
and
$$\displaystyle
\Mdyn  M_{gas}  = (\alpha\,\LprimeCO)^{2}
\ \ ,
\eqno(6)$$
where $\nbarH2$ is the gas  density averaged over the whole volume.
  The quantity $\alpha \LprimeCO$
measures the geometric mean of total mass and gas mass.  It
underestimates total mass and overestimates gas mass.
Hence if the CO emission in ULIRGs comes from
regions not confined by self-gravity, but instead from an intercloud
medium bound by the potential of the galaxy, or from molecular gas
in pressure, rather than gravitational equilibrium, then the usual  
relation
$M_{\rm gas}/$\LprimeCO $=\alpha$ must be changed. The effective $\alpha$ is
lower than $2.6n^{1/2}/T_{\rm b}$.

Extensive high-resolution mapping of CO emission from ULIRGs shows that 
  the molecular gas is in
rotating disks or rings. Kinematic models (Downes \& Solomon 1998) in which most 
  of the CO flux comes
from a moderate density warm intercloud medium  have been used to  
account for the rotation
curves, density distribution, size, turbulent velocity, and mass of  
these molecular rings. Gas
masses were derived from a model of radiative transfer rather than the use  
of a standard conversion
factor. The models  yield gas masses of $\sim5 \times 10^{9} $\,  
\Msun  ,  approximately five times lower
than the standard method, and a ratio $M_{\rm gas}/L^\prime_{\rm CO}$
$\approx$ 0.8\,\Msun\ (K\,\kms\,pc$^2)^{-1}$.   The ratio of gas  to  
dynamical mass $M_{\rm
gas}/M_{\rm dyn}\approx$ 1/6 and a maximum ratio of gas to total mass  
surface density
$\mu/\mu_{\rm tot}= 1/3$.  This effective conversion factor $\alpha = $ 
  0.8\,\Msun\
(K\,\kms\,pc$^2)^{-1}$ for ULIRGs has been adopted for EMGs by many  
observers of high-$z$ CO
emission  and we use it throughout this review. However, until a  
significant number of EMGs are
observed with sufficient angular resolution to enable a calibration of  
$\alpha$, the extrapolation
in the use of $\alpha = 0.8$ to EMGs from ULIRGs must be regarded as  
tentative.

\subsection{Classification of the EMGs}

The list of 36 EMGs reported in the literature at the time of this  
review are given in Table 1, together with their derived properties.  
The gas masses  were calculated using the luminosity of the lowest available CO transition and $\alpha=0.8$ (see Section  
2.2).  All quantities assume the cosmology adopted for this review.  
Appendices 1, 2, and 3 at the end of this article give the 
observed properties from which the quantities in Table 1 were 
calculated.\footnote{Appendix 1 lists 
coordinates, redshift, galaxy type and magnification for each EMG.  
Appendix 2 gives velocity integrated flux densities ($S\Delta v$), linewidths as full width at half-maximum (FWHM) ($\Delta v$),
peak line flux densities ($S$),  line luminosities ($L'$) for the CO  
transitions observed in the
EMGs, and inferred molecular gas masses.  The observed quantities  
listed are those reported in
the references cited, after adjustment for the cosmology assumed in  
this review.  Where lens
models exist, intrinsic luminosities are listed, calculated using the  
magnifications given in
Appendix 1.   In addition to CO,
data for detections of HCN are listed, as well as  for CI whose fine-structure lines  originate
from interstellar molecular gas.  Appendix 3  gives the observed continuum 
flux densities at various  wavelengths of the EMGs,
together with the inferred FIR luminosity, including the intrinsic  
luminosity where it is possible
to correct for lens magnification.   Brackets indicate the measurements 
  that were included in
the calculation of the listed luminosity values cited.  Frequently,  
only a single measurement is
used to estimate the luminosity, together with a set of assumptions, so 
  the values listed
should be regarded with caution.}  The overwhelming majority of these  
detections were made with the IRAM interferometer.   
Lists of EMGs have been constructed by Cox et al. (2002), Carilli et al.  
(2004), Hainline et al. (2004), and Beelen (http://www.astro.uni-bonn.de/$\sim$beelen/database.xml).  The sources are  
listed in all tables and appendices in order of redshift.  No blind survey for high-$z$ 
CO emission has been done because of its prohibitive cost in observing time  
with present instruments.  Were such a blind survey to be done  
eventually by ALMA, it could  
result in additional types of EMGs.  Figure 1 shows the number of EMGs by type  
as a function of redshift.  Despite the selection effects that attend  
the detection of EMGs, one can see that the current flux-limited sample broadly reflects the epoch where most star  
formation in the Universe is currently thought to occur.

With recent improvements in
millimeter bolometers, large numbers of quasi-stellar objects (QSOs) have been observed in 1.2-mm continuum emission.
Approximately 30\% of the bright QSOs at all redshifts $z > 2$ are   strong millimeter/submillimeter continuum
emitters with a typical inferred rest-frame luminosity of $\lfir \sim  
10^{13}$ \lsun\ (Izaak et al 2002, Omont et al. 1996a).  The percentage of  
submillimeter detections is higher (60\%) for gravitationally lensed  
quasars (Barvainis \& Ivison 2002).
Identifying the redshift  appropriate for a CO emission search can be
difficult because the molecular gas in the host galaxy may have a
significantly different redshift from the broad optical emission line  
region of the QSO.
A key question for the EMGs identified with QSOs is
whether the FIR luminosity is powered by rapid star formation
(starbursts) in the molecular clouds or by the active galactic nucleus  
(AGN) that may be accreting
molecular gas.

In SMGs, unlike the optically selected  
quasars, the total luminosity is
completely dominated by their (rest-frame) FIR  emission.  The surveys  
at 850 \microns ,
primarily carried out with the Submillimetre Common-User  Bolometer  
Array (SCUBA)
instrument on the James Clerk Maxwell Telescope (JCMT)  have found   several hundred galaxies, or about 1  arcmin$^{-1}$ (see, for example, Scott et al. 2002). They
represent a substantial part of the FIR background and may contribute  
as much as half of all
star formation at high $z$.  Although many SCUBA galaxies harbor active  
galactic nuclei (AGNs), the
AGNs contribute only a small fraction of the bolometric luminosity,  
which is dominated by star
formation (Alexander et al. 2004). Only a small subset of about 15  
blank-field submillimeter
sources have been observed in CO emission.

A relatively small proportion (19\%) of EMGs are identified with radio  
galaxies. Radio galaxies
are a rare population and are not selected for being gravitationally  
lensed.  However, seven
IR-luminous radio galaxies have been observed in CO emission, and these 
  include some of the
more interesting examples.

The identification of a set of EMGs with LBGs  
would be significant in that
it would tie the EMGs to a huge population of early Universe objects.   
However, only a single LBG
has been detected in CO emission (Baker et al. 2004).  The low CO line  
luminosity of this object
compared with the other EMGs suggests that LBGs form a different  
class of early Universe
galaxies, something that remains to be confirmed using ALMA.

\subsection{Examples of EMGs}

This section presents and discusses EMGs by type and historically within each type.

\subsubsection{IRAS F10214}

In 1991, IRAS FSC10214+4724 was shown to be an extraordinarily luminous 
  high-redshift
IR source  (Rowan-Robinson et al. 1991). With a redshift of $z=2.3$ it 
  was by far the most
luminous IR galaxy yet found, more than 30 times as luminous as  
local ULIRGs.  Shortly
after IRAS F10214 was identified, the first high-$z$ CO emission was  
searched for and found in
the (3--2), (4--3), and (6--5) lines (Brown \& Vanden Bout 1991, 1992; Solomon, Downes \& Radford 1992a).  Allowing
for the negative K-correction, Solomon, Downes \& Radford (1992b) found the CO line  
luminosity, $L^{\prime}_{\rm CO}$, calculated
from the flux measured at the IRAM 30m Telescope, to be 100 times less than first estimated, but still about  
an order of
magnitude greater than that in any galaxy in the local Universe, yielding a  
molecular gas mass of
10$^{11}$ \Msun,  equal to the baryonic mass of an entire  large  
galaxy.  (Agreement between the 12-m and 30-m measured fluxes was obtained with new observations at the 12-m by Radford et al. (1996)).  The strong CO(6--5)
line, originating from a rotational level 116 K above the ground state,  and the (6--5)/(3--2)
line ratio indicates the presence of moderately dense gas substantially warmer than most of
the molecular mass in Milky Way GMCs or normal spiral galaxies.

Optical and near-IR spectroscopy show both
narrow and broad emission line systems, with the narrow lines  
indicating a Seyfert 2
nucleus (Lawrence et al. 1993)  and the  broad lines observed in polarized  
light indicating
the presence of an obscured quasar (Goodrich et al. 1996).

High-resolution optical and near-IR imaging (Broadhurst \& Leh'ar 1995, Graham \& Liu 1995, Matthews et al. 1994)
clearly show that F10214 is gravitationally lensed.  The 2.2-\micron\  
image shows a compact
0.7$\arcsec$ diameter source superposed on a weaker 1.5$\arcsec$ arc.   
CO maps of the (6--5)
line with the IRAM interferometer show an elongated structure that was 
  modeled as a CO arc
convolved with the interferometer beam and fit to the CO data  
(Downes, Solomon \& Radford 1995). From the
length of the CO arc, the apparent CO luminosity, the linewidth, and  
the intrinsic brightness
temperature of the line (deduced from line ratios), Downes, Solomon \& Radford (1995) 
  derived a magnification
    $\mu = 10 f_{\rm v}$, where $f_{\rm v}$ is the velocity filling factor, or fraction of the full line width intercepted by a typical line of sight.  This  
magnification reduced the
intrinsic CO line luminosity and molecular mass to that of local  
ULIRGs. The radius of the
molecular ring was found to be  600$/ f_{\rm v}$ pc, much larger than that of the AGN 
  torus and similar to
that in ULIRGs, but much less than that of a full galactic disk. The  
magnification for the FIR radation
was 13, and for the mid-IR it was 50.

Recent improved high-resolution maps of CO(3--2), (6--5), and  
(7--6) (Downes \& Solomon, manuscript in preparation) show
that the size of the lensed CO image is 1.6$\arcsec \times \leq  
0.3\arcsec\ (2.7 \times \leq0.5$ kpc). More importantly, a
velocity gradient is observed along the arc and line profiles
  show two distinct kinematic components at the east and west sides,  
demonstrating that the
molecular emission originates in a rotating disk around the quasar.  
Positions, sizes, and
linewidths are the same in all three lines, indicating that they  
originate in the same volume with
the same kinematic distribution. The line ratios indicate a mean  
emission-weighted kinetic
temperature of 50 K and a mean \Htwo\ density of 3000 cm$^{-3}$. A  
search for $^{13}$CO
emission yields a ratio of $^{12}$CO/$^{13}$CO $\geq$ 21, which is similar to  
high values found in ULIRGs
but higher than those of nearby spiral galaxies, indicating a modest  
opacity for   $^{12}$CO. The
true size of the molecular ring, the CO luminosity, molecular mass, and 
  the excitation of the CO
ladder all look similar to those observed in local ULIRGs.

Vanden Bout, Solomon \& Maddalena (2004) observed strong HCN(1--0) emission from F10214 with an  intrinsic  line luminosity
similar to that in local ULIRGs such as Mrk 231 and Arp 220. HCN  
emission traces dense gas
generally associated with the star-forming cores of GMCs (see Section  
3.1).  The very high ratio of HCN to CO luminosities
$\lcop$/$\Lhcnp = 0.18$ is characteristic of starbursts in the local  
Universe. All galaxies with global HCN/CO
luminosity ratios greater than 0.07 were found to be luminous
 ($\Lfir > 10^{11}$ \lsun) starbursts (Gao \& Solomon 2004).  F10214  contains both a dust-enshrouded quasar responsible for the mid-IR luminosity and a much larger molecular ring starburst responsible for a substantial fraction of the FIR luminosity.

\subsubsection{Cloverleaf}

Hazard et al. (1984) found the quasar H1413+1143 (better known as the Cloverleaf), a broad  
absorption
line QSO at a redshift of $z=2.55$. It  
was subsequently identified
optically as a lensed object with four bright image components  
(Magain et al. 1988). Barvainis, Antonucci \& Coleman (1992) discovered strong FIR and submillimeter radiation from the  
Cloverleaf, indicating a
substantial dust component  with a FIR spectral energy
distribution (SED) similar to that of IRAS F10214. This was the first  
indication that some
bright optical high-$z$ quasars also are extremely  IR luminous.

Redshifted strong CO(3--2) emission was observed using both the IRAM  
30-m Telescope and
  Plateau de Bure Interferometer (Barvainis et al. 1994) with an apparent  
line luminosity
about three times greater than that from F10214. Barvainis et al. (1997) observed three additional  
rotational lines (4--3), (5--4), and
(7--6) were observed  at the IRAM 30m Telescope and  
their line ratios
used to constrain the physical conditions of the gas and the CO to  
\Htwo\ conversion factor.
These measurements showed $\Lcop(4-3)>\Lcop(3-2)$, indicating a high  
kinetic temperature
and low optical depths.   More recent measurements (Wei\ss\ et al. 2003) show  
a higher (3--2) flux
and a lower line ratio (4--3)/(3--2) indicative of lower kinetic  
temperatures and  subthermal
excitation.
  The Cloverleaf CO emission lines have a higher flux density than do the lines 
  from  any other high-$z$ source, owing to both powerful intrinsic line luminosities and magnification. As 
  a result,
they can be successfully imaged at high angular resolution. The lensing 
  also magnifies the scale
of the emission making it possible to deduce true source size at scales 
  below the instrumental
resolution.

  Using the millimeter array at the Owens Valley Radio Observatory (OVRO),
Yun et al. (1997) obtained an interferometric map of the Cloverleaf in  
which the
CO(7--6) emission was partially resolved. They used Hubble Space  
Telescope (HST) images to model a lens with an
elliptical potential and an external sheer. This model constrained the  
intrinsic size of the
  CO(7--6) source, which has a radius of approximately 1100 pc.  
Separation of the red and
blue line wings showed a kinematic structure consistent with a rotating 
  disk.
Alloin et al. (1997) obtained a high-resolution map (0.5$''$)  with the 
  IRAM interferometer
that clearly resolved the emission into four spots  similar to the  
lensed optical radiation.  Figure 2 shows an image of the CO(7--6)  
emission  contructed by Venturini \& Solomon (2003)from their data. A
model based on  HST and  Very Large Array (VLA) images gave an upper limit 
  to the source radius of approximately
1200 pc. Kneib et al. (1998) used enhanced IRAM CO(7--6) 
images and HST images to
construct two lens models using a truncated elliptical mass  
distribution with an external
shear (galaxy + cluster). From the separation of the kinematic  
components and the HST-based lens model they deduced a CO radius of only 100 pc and a  
magnification of 30. This size
scale is characteristic of an AGN torus.

Venturini and Solomon (2003)  fit a two-galaxy lensing model directly  
to the IRAM CO(7--6) map
rather than to the optical HST image. The fit  obtained by minimizing  
the difference between
the map produced by the lensed model  and the IRAM CO(7--6) image   
yielded a source with
disklike structure  and a characteristic radius of 800 ${\rm pc}$, a  
value similar to that of the
CO-emitting regions present in nearby starburst ULIRGs. The model
reproduces the geometry as well as the brightness of the four images of 
  the lensed quasar. The
large size of the CO source seems to rule out a scenario in which the  
molecular gas is
concentrated in a very small region around the central AGN. With the  
magnification of 11 found
from this model and the CO(3--2) flux given by Wei\ss\ et al. (2003),   
the total molecular mass
is $3.2\times10^{10}$
\Msun, with a molecular surface density of 10$^4$ \Msun\ pc$^{-2}$.  
Wei\ss\ et al. (2003)
argue that using $\Lcop$(3--2) rather than $\Lcop$(1--0) has only a  
10\% effect on the
calculated molecular mass.  The dynamical mass of the rotating disk is  
$\Mdyn sin^{2}i=2.5\times
10^{10}$ \Msun.

HCN emission traces dense gas
generally associated
with the star-forming cores of GMCs.
  Strong HCN(1--0) emission has  been observed from the
Cloverleaf (Solomon et al. 2003) with an intrinsic line luminosity slightly  
higher than that in local
ULIRGs, such as Mrk 231 and Arp 220, and 100 times greater than that of 
  the Milky Way.  To put
this in perspective, the intrinsic HCN luminosity of the Cloverleaf is  
10 times greater than the CO
luminosity of the Milky Way, indicating the presence of 10$^{10}$  
\Msun\ of dense
star-forming molecular gas.

The molecular and IR luminosities for the Cloverleaf show that the
large mass of dense molecular gas indicated by the HCN luminosity could 
  account for a
substantial fraction (from star formation), but not all, of the IR  
luminosity from this quasar.
If Arp 220 is used as a standard for the luminosity ratio \Lfir\ /$L^{\prime}_{\rm HCN}$,   
star formation in the dense
molecular gas could account for $5\times10^{12}$ \Lsun\ ,  or about 
20\%  of the total intrinsic IR
luminosity.   Using the highest ratio for a ULIRG gives an upper limit
of 40\%.  

The  model by (Wei\ss\ et al. (2003)
  of the IR spectral energy distribution of the Cloverleaf has two
distinct components: one with a warm dust temperature $T_{\rm d}=115$ K  
responsible for the mid-IR,
and the other much more massive component with $T_{\rm d}=50$ K that produces 
  the FIR.  The model
FIR luminosity, 22\% of the total, may correspond to the luminosity  
generated by star
formation and the mid-IR to heating by the AGN.  Using the model \Lfir\ yields  \Lfir\  
/\Lhcnp=1700,  comparable
to that of ULIRGs and only a factor of 2 higher than that for normal spiral
galaxies (Gao \& Solomon (2004). The star formation rate per solar mass of dense  
gas  is then
similar to that in ULIRGs and only slightly higher than that in normal  
spirals.

\subsubsection{VCV J1409+5628} This EMG is an optically luminous  
radio-quiet quasar with the
strongest 1.2-mm flux density found in the survey by Omont et al. (2003).  It 
  has been observed in both
CO(3--2) and CO(7--6) emission (Beelen et al. 2004).  The line luminosity  
of $\Lcop (app.) = (7.9\pm
0.7) \times 10^{10}$ \Kkmspc\ leads to a gas mass of $M_{\rm gas} = 6.3  
\times 10^{10}\mu^{-1}$ \Msun,
which is $\sim$20\% of \Mdyn\ for reasonable inclinations.  If the  
extent of the radio continuum,
from a VLA image at 1.4 GHz, represents the extent of the CO emission,  
the molecular gas is
confined to a torus or disk of diameter 1--5 kpc.  This is similar both to  
the molecular gas extents
inferred from lens models of F10214 and the Cloverleaf and to what is  
observed in ULIRGs.

\subsubsection{PSS J2322+1944} This EMG is an IR-luminous quasar.  
The extent of its
molecular gas has  been inferred from a remarkable gravitationally  
lensed image of the CO
emission --- a so-called Einstein Ring.  
  Carilli et al. (2003) studied this lensed system on
sub-kiloparsec scales with the $0.6\arcsec$ resolution of the VLA at 43 
  GHz, where the CO(2--1)
line from this $z=4.12$ object is redshifted. The VLA image is shown in 
  Figure 3.  The data are
consistent with a dynamical mass of $\Mdyn = 3\times10^{10}sin^{-2}i$ 
\Msun\ and
confinement of the molecular gas in a  disk of diameter 2.2 kpc.    The 
radio continuum is  co-spatial with the
molecular gas and the star formation rate is $\sim$900 \Msun\  
year$^{-1}$.  PSS J2322+1944 is the
fourth EMG to be observed in \fCI emission.  This object provides  
strong evidence for the
presence of active star formation in the host galaxy of a luminous  
high-redshift quasar.

\subsubsection{BR 1202-0725} This is an optically bright radio-quiet  
quasar,  the third EMG to be
discovered (Omont et al. 1996b), and the first to show multiple components.  
  Whether these two
components, separated by 4$\arcsec$, are companion objects or the  
result of gravitational
lensing remains an issue.  High-resolution imaging Carilli et al. 2002a)  
using the VLA of the CO(2--1)
emission has shown that the
southern component is roughly twice as massive as the northern  
component, and there is a
significant difference in the velocity widths of the CO lines of the  
two components.  This finding provides evidence
against the presence of a gravitational lens.  However, the total  
molecular gas mass exceeds the
dynamical mass of the system unless an unreasonably low value of   
$\alpha$   is used to
calculate $M_{\rm gas}$.  Magnification by a gravitational lens would allow 
  for more reasonable
values of $\alpha$.

\subsubsection{APM 08279+5245} This extremely luminous broad absorption  
line quasar was accidently discovered in a survey for cool carbon  
stars (Irwin et al. 1998).  The high redshift of
$z=3.9$ would have made it the most luminous known object in the  
Universe were it not for the
magnification of a gravitational lens (Egami et al. 2000).  The  
magnification at optical wavelengths
can be as large as $\mu=100$; for CO emission it is much less, $\mu=7$  (Downes et al. 1999, Lewis et al. 2002).  The CO (4-3) and (9-8) emission was first observed in APM08279  
with the IRAM
interferometer
Downes et al. 1999).  The strong (9--8) emission indicates the presence of  
hot dense gas with a kinetic
temperature of approximately 200 K. The observed ratio of \Lfir/\Lcop   
is twice  that of other
EMGs.  In addition to the central  molecular emission region,
observed in four CO transitions, high-resolution images of the CO(2--1) 
  emission with the VLA
reveal two emission regions lying to the north and northeast,  
2--3$^{\prime\prime}$ distant from the
central region (Papadopoulos et al. 2001).  If real, these could be  
companion galaxies.  The nuclear
CO(1--0) emission is imaged in a (partial) Einstein Ring (Lewis et al. 2002).

\subsubsection{SDSS J1148+5251} This is the most distant known quasar,  
with a redshift of
$z=6.42$.  It was shown to be an EMG via the observations of CO(3--2)  
emission using the VLA,
and CO(6--5) and CO(7--6) emission using the IRAM interferometer  
(Bertoldi et al. 2003b, Walter et al. 2003).  The CO observations imply a mass of molecular gas  
$M_{\rm gas} = 2.1 \times
10^{10}\mu^{-1}$ \Msun.  The thermal dust emission (Bertoldi et al. 2003a)
leads to  a star formation rate of
$\sim3000\mu^{-1}$ \Msun\ year$^{-1}$.  This is clear evidence for the presence  
of vast amounts of
molecular gas, composed of heavy elements, only $\sim$850 million years 
  following the Big
Bang.  High-resolution ($0.17^{\prime\prime} \times  
0.13^{\prime\prime}$; $\leq$ 1 kpc)
imaging of the CO(3--2) emission using the VLA (Walter et al. 2004), shown  
in Figure 4, suggest that
this source may be a merger of two galaxies.

\subsubsection{SMM J02399-0136} This SMG was the first SCUBA source  
identified as an EMG
(Frayer et al. 1999), using OVRO.  It is the brightest galaxy detected in an early SCUBA 
  survey of rich lensing
clusters (Smail, Ivison \& Blain 1997).  J02399  harbors an AGN  
(Ivison et al. 1998).  The observed
integrated line strength of the CO(3--2) line, with the observed CO  
redshift of $z=2.808$, leads to
$\Lcop (app) = 12\times10^{10}$ \Kkmspc.  Correction for a cluster  
lens magnification of
$\mu=2.5$ yields $\Lcop (int) = 4.9\times10^{10}$ \Kkmspc.  This is  
comparable to CO luminosities
for ULIRGs, and was the first evidence that  SCUBA sources identified  
as EMGs may be similar in
nature to ULIRGs. Higher resolution observations of the CO emission at  
IRAM confirmed the OVRO
detection (Genzel et al. 2003).  These data were fitted to a rotating disk  
model very similar but
larger in size than that seen in ULIRGs: a molecular gas mass $M_{\rm gas}  
= 3.9\times 10^{10}$
\Msun\ confined within a radius of 8 kpc.  This source remains one  
of few EMGs with the
potential for molecular gas to be extended in a disk with radius larger 
  than 2 kpc.

\subsubsection{SMM J14011+0252} This SMG was the second SCUBA source  
from the Lensing
Cluster Survey (Smail, Ivison \& Blain 1997) to be detected in CO emission; it has  
been heavily observed since
being identified as an EMG.  There is no evidence  for the presence  of 
an AGN in J14011.  The
detection of CO(3--2) emission (Frayer et al. 1999) at OVRO was followed by  
more interferometry to
determine the location of the CO source among the 850-$\mu $m peaks in  
the SCUBA image and
its extent.  From combined OVRO and Berkeley-Illinois-Maryland Association (BIMA) observations it was argued  
(Ivison et al. 2001) that the
CO emission was extended on a scale of diameter 20 kpc, assuming a cluster  
magnification of
$\mu=2.5$, well beyond what is seen in ULRIGs.  Higher signal-to-noise  
observations at IRAM
(Downes \& Solomon 2003) did not confirm this extent, as the CO emission is  
confined to an observed disk of only $2.2\arcsec$, or a 
diameter $\leq7$ kpc for a magnification of 2.5.  

\subsubsection{SMM16359+6612} This is a somewhat lower luminosity  
($\lfir=10^{12}$ \lsun)
SMG that nevertheless has been observed in CO(3--2) emission  aided by  
a gravitational lens
that provides a total magnification factor of $\mu=45$.  The image 
obtained  with the IRAM
Interferometer
(Kneib et al. 2005a), together with spectra of the three image components,  
is shown in Figure 5. CO
observations of SMM J16359 have also been reported by Sheth et al.  
(2004).  This is the third
SMG reported to have spatially resolved CO emission.  Here, the quality 
  of the data together
with the lens model of Kneib et al. (2004b) leads to an inferred disk  
size of $3\times1.5$ kpc.
Whereas the FIR luminosity is comparable to that of Arp 220, the CO  
luminosity is approximately half that
of Arp 220.  The mass inferred from the CO luminosity is 30\% or 60\% of the  
calculated dynamical
mass for a ring-disk structure or a merger, respectively.

\subsubsection{4C41.17}  This is one of only seven radio galaxies to be observed in CO emission.  High-$z$ radio galaxies (HzRGs) have been difficult to detect 
  in CO emission because
the candidates searched are not gravitationally lensed and the observed peak 
  CO flux densities are
small ($\sim2$ mJy).  Stevens et al. (2003) have argued that HzRGs  
and their companions,
revealed in deep 850-$\mu$m images,  form central cluster  
ellipticals.  Four of the seven
HzRG examples cited by Stevens et al. (2003), including 4C41.17,  are also EMGs.   A
position--velocity plot of the   CO(4--3) emission   (De Breuck et al. 2005), 
clearly reveals
two components.  Both are gas-rich systems, each with  $M_{\rm gas} \sim 3 
\times  10^{10}$ \Msun.
Their velocity separation leads to a dynamical mass $\Mdyn \sim6 \times 
  10^{11}sin^{-2}i$ \Msun, for
the potential binding the components.  The system could be two gas-rich 
  galaxies merging to
form a massive cD elliptical galaxy.

\section{DISCUSSION}

\subsection{Molecular Gas Mass and Star Formation Efficiency}

The intrinsic line luminosities  given in Table 1 have been corrected  
for magnification  for
those sources with known lensing and published estimates of the   
magnification. For
sources without apparent lensing we have adopted the measured line 
luminosity (assumed
the magnification
$\mu=1$) in the figures and discussion of this section. The  CO line  
luminosity of EMGs covers
a wide range of
$\LprimeCO = (0.3-16) \times 10^{10}$ \Kkmspc. Not surprisingly, because this is basically a flux-limited sample, the  
lowest line luminosities
occur for sources (primarily QSOs) with high magnification. The average CO line luminosity is $\langle$log (\Lprimeco)$\rangle$ 
=  10.45$\pm 0.47$
corresponding to an average gas mass of $2.3 \times 10^{10}$ \Msun\  
using $\alpha=0.8$.
There is little difference between the average CO luminosities among the  
three categories of
sources QSOs, SMGs, and radio galaxies.

Figure 6 shows the CO line luminosity (for the lowest $J$ transition for 
which  data exist) as a
function of redshift for  EMGs and samples of ULIRGs, luminous IR galaxies (LIRGs), and normal 
spirals. In
comparison with EMGs the average line  luminosity for ULIRGs in the 
local Universe is smaller
by about a factor of 3 and with a much  smaller range,   log 
$(\Lprimeco) = 9.98
\pm 0.13$ (Solomon et al. 1997) . However, there is significant overlap 
between CO luminosities  from
these high-$z$ galaxies and those in the local Universe including 
ULIRGs, LIRGs, and  even some
normal spirals. For example, the ULIRG 20087-0308 has a CO line 
luminosity of  1.8
$\times 10^{10}$
\Kkmspc,  larger than that of approximately one third of the EMGs. Local interacting galaxies 
  with much more
modest IR luminosities such as Arp 302 also have CO luminosities close  
to the midrange found
in EMGs. The normal, isolated spiral NGC3147 has a CO luminosity of $0.7  
\times 10^{10}$
\Kkmspc, larger than six of the EMGs. Most normal, large spiral 
galaxies  have a CO luminosity
about a factor of 5--10 less than that of ULIRGs and  10--30 times less than that of  
EMGs.

Assuming a constant conversion factor, EMGs have on average a higher  
molecular gas mass
than the most gas-rich local Universe galaxies, but only a few times  
higher.  In the local
Universe there appears to be a ``ceiling'' for ULIRGs with  $M_{\rm gas} <  
  2 \times 10^{10}$
\Msun. Approximately two-thirds of the EMGs lie above this local maximum with a 
typical gas  mass of $5
\times 10^{10}$ \Msun. (about 30 times the molecular mass of the Milky 
Way)  This difference
between local and high-redshift gas masses may be important in 
understanding   the nature
of the high-$z$ galaxies and early galaxy evolution. One possibility is 
that  EMGs have the
same molecular gas mass as do ULIRGs but have a lower CO to \Htwo
\ conversion factor. Or, they may have the same conversion factor and thus  contain more
molecular mass, possibly distributed over a larger disk. We assume the  
conversion factor is
the same here and in the following sections.

The ratio of  FIR luminosity to CO luminosity, \Lfir /\Lprimeco\ is an  
indicator of the star
formation rate per solar mass of molecular gas and is often taken as a  
measure of the star
formation efficiency (Young, et al. 1986, Solomon \& Sage 1988). Figure 7 shows this ratio as a function of  
redshift.  The star formation
efficiency for the EMGs at high $z$ is similar to or slightly higher  
than that for ULIRGs in the local
Universe with an (logarithmic) average
$\Lfir /\Lprimeco = 350$; this translates into a star formation  
efficiency $\Lfir /M_{\rm gas} =
430$ \Lsun/\Msun.

It is well known (Sanders et al. 1988, Sanders \& Mirabel 1996, Solomon \& Sage 1988) that the star  
formation efficiency of
ULIRGS, which are mergers and closely interacting galaxies, is higher  
than that of normal
spiral galaxies and there is a well-established trend whereby star 
formation  efficiency  increases
with increasing FIR luminosity.  Figure 8, which shows  log$(\Lfir)$\  as 
a  function of
log$(\Lprimeco)$\ for normal spirals, LIRGs, ULIRGs, and EMGs, extends 
the trend above
10$^{13}$ \Lsun. The slope is 1.7, similar to that found without EMGs
(Gao \& Solomon 2004).  This demonstrates that, given their high FIR luminosity, EMGs have  the high star formation  
efficiency  expected by extrapolation from low-redshift galaxies. Figure 8 also shows  
that EMGs with the
same CO luminosity or molecular mass as ULIRGs also have the same (or  
slightly higher) FIR
luminosity and star formation efficiency. They do not look like scaled-up versions of normal
spirals with a larger molecular mass.  The high star formation  
efficiency of luminous IR
galaxies is due to a very high fraction of dense molecular gas as traced  
by HCN emission
(Solomon, Downes \& Radford 1992c) and other molecules, rather than the total molecular 
gas mass traced by
CO  emission. In this sense, CO luminosity is not a linear tracer of 
the star formation rate.

\subsection{Star Formation and Gas Depletion Lifetime}

The high star formation efficiency of EMGs also implies a short star 
formation lifetime. Taking the
star formation rate to be given by 1.5 x $10^{-10}\Lfir$\   [\Msun\ 
year$^{-1}$], see for example
Kennicutt(1998), and using the above star formation  efficiency $\Lfir 
/\Lprimeco = 350$ and
$\alpha = 0.8$, the average star formation rate per solar mass of 
molecular gas   $\approx 6
\times 10^{-8}$ year$^{-1}$. (This assumes that all FIR luminosity 
is due to star formation) The
inverse is the average star formation lifetime\,  or average gas depletion time
$\tau_{\rm SF} = 16$ My. Starbursts in EMGs are a  brief but critical  
phase in galaxy formation and
evolution.

Figure 9 shows the star formation lifetime of normal spirals, ULIRGs, 
and EMGs as a function of FIR
luminosity. Because the  mass conversion factor of CO to \Htwo\ is larger for 
normal spirals than for ULIRGs,
$\alpha$ is treated as a parameter and the lifetime is normalized to 
$\alpha = 1$.  For normal
spirals $\alpha = 4.6$ and the gas lifetime will be larger than 
indicated. For ULIRGs and, presumably,
EMGs,  the lifetime is close to that indicated. Normal spirals with  
dust-enshrouded star
formation have gas depletion times in excess of $10^9$ years, whereas 
ULIRGs and EMGS have lifetimes
in the range   $10^7$ to  $10^8$ years.  For EMGs the lowest level CO 
line observed
has been used to determine the molecular gas mass; to the extent that 
the CO (1--0) line luminosity
is higher than the (3--2) or (4--3) line the gas mass and  lifetime will 
be proportionally larger
for some EMGs without CO(1--0) measurements. The few available (1--0) 
measurements indicate
that this will be a small effect (less than a factor of 2) for most 
sources.(This short lifetime also
sets limits on the dimensions of the starburst because the dynamical time 
must be less than the
starburst lifetime.)

\subsection{HCN, [CI], \&\ [CII] Emission}

\subsubsection{Hydrogen Cyanide: Dense Molecular Gas}

HCN emission traces dense gas, $n(\Htwo) > 3\times10^{4}$\ cm$^{-3}$  
generally associated
with the star-forming cores of GMCs, whereas CO, with its low dipole  
moment,  can have
emission excited by gas at the much lower densities found in GMC 
envelopes. HCN  line
luminosity is a much more specific tracer of star formation than CO 
luminosity,  although CO
is a better overall tracer of total molecular mass.  In normal spirals 
and  luminous  infrared
galaxies (LIRGs and ULIRGs),  the correlation between FIR luminosity 
and HCN line  luminosity
is much tighter than that of FIR with CO line luminosity  
(Gao \& Solomon 2004; Solomon, Downes \& Radford 1992c). The star
formation rate deduced from the IR luminosity scales linearly with the 
amount of dense
molecular gas traced by HCN emission over more than three orders of 
magnitude in IR
luminosity from  $ 10^{9.3}$ to
$10^{12.3}$ \Lsun. This is not the case for CO emission which shows   
much higher  star
formation efficiencies, indicated by \lfir /\Lcop, for luminous IR  
galaxies than for normal
galaxies. In particular, ULIRGs have a star formation efficiency
or rate of star formation per solar mass of molecular gas that is, on average, a five times higher  
 than that of normal  
galaxies. Luminous IR
galaxies have a huge HCN line luminosity, large mass of dense  gas, and 
a high ratio of dense
gas to total molecular gas indicated by  \Lhcnp/\Lcop; for  ULIRGs 
this luminosity ratio is typically 1/4
to 1/8, whereas for normal spirals it is in the range 1/25 to 1/40.   The 
ULIRG Mrk 231 often
regarded primarily as an AGN has a ratio \Lhcnp/\Lcop = 1/4 and an  
HCN luminosity much
larger than the CO
  luminosity of the Milky Way. This finding led Solomon, Downes, \& Radford (1992c) to  
conclude that even this galaxy
with a definite AGN had most of its bolometric luminosity supplied by   
a starburst. This has
recently been confirmed by  near-IR spectroscopy of the Mrk 231  
starburst disk
(Davies, Taconi \& Genzel 2004). All galaxies in the local Universe with global ratios
\Lhcnp/\Lcop $\geq
$ 1/14  are luminous or ultraluminous IR  starburst galaxies  
(Gao \& Solomon 2004).

HCN observations of EMGs provide an important test of the star  
formation model. The fact
that the ratio of IR luminosity to HCN luminosity  in ULIRGs 
is the same as  in lower luminosity
normal spiral galaxies shows that ULIRGs, like the lower luminosity  
galaxies, are primarily
powered by star formation and that the HCN line luminosity is a good  
measure of the mass of
actively star-forming cloud cores
(Gao \& Solomon 2004; Solomon, Downes \& Radford 1992b).  The star formation that is responsible for  
the FIR emission has a
rate that is linearly proportional to the HCN luminosity tracing the  
mass of dense molecular
gas but not to the total molecular gas as traced by CO.
  HCN observations can address the question of whether EMGs have a  
sufficient mass of
dense molecular gas to account for the huge IR luminosity by star 
formation.

HCN(1--0) emission has been detected from three EMG: the  
Cloverleaf
(Solomon et al. 2003), F10214
(Vanden Bout, Solomon \& Maddalena 2004), and VCV J1409 (Carilli et al. 2004). In all three  
cases, the HCN(1--0) line
luminosity is larger by a factor of 100  (or more) than that of normal  
spiral galaxies and a few
times that of the ULIRG Arp 220, indicating the presence of a large  
mass of dense molecular
gas. Based on the FIR luminosity (not the mid-IR from very hot dust) the  
ratios  \Lfir/\Lhcnp\ =
1700 and 2700 for the Cloverleaf and F10214, respectively, are only  
slightly higher than that of
Arp 220 or the average for local ULIRGs.  The  dense gas fraction  
indicators \Lhcnp/\Lcop =
1/14 and 1/6, respectively, denote starbursts in both systems.  
Detailed  discussions of
the HCN in these two objects are given in Section 2.4. The third 
detection VCV J1409 shows not only
the  highest HCN luminosity (assuming no magnification by a 
gravitational lens) but also a
somewhat   higher
\Lfir/\Lhcnp\ = 4000, approximately a factor of 3 above the average for local  
IR starbursts. Using a
dense gas conversion factor for the HCN luminosity $\alpha_{\rm HCN} \approx$  
7 \Msun\
(\kkmspc)$^{-1}$  (Gao \& Solomon 2004) leads to a dense gas mass of 1, 4, and  
5$\times10^{10}$
\Msun\ for F10214, the Cloverleaf, and VCV J1409, respectively, where  
the mass of dense gas
in VCV J1409 assumes no magnification by a gravitational lens. Assuming 
  that all of the FIR
luminosity is from star formation  leads to lifetimes for the dense gas 
  of approximately
10--20 million years.

There are four other EMGs with upper limits for HCN (Carilli et al. 2004; Izaak et al. 2002); all seven
high-$z$ sources including the upper limits are within the range 
expected from  an extension
of the low-$z$ galaxy FIR-HCN linear correlation if star formation is  
responsible for most of
the FIR luminosity (Carilli et al. 2004).

\subsubsection {Atomic Carbon}

Observations of the forbidden fine-structure lines of neutral atomic  
carbon in the Milky Way
and nearby galaxies (Ojha et al. 2001, Gerin \& Phillips 2000, and  
references therein) have revealed
a close association with CO emission.  Because the critical density for 
  excitation of  both
the  \fCI(\CIl $\rightarrow$ \CIll) transition at 492.160  
GHz and the (\CIu
$\rightarrow$  \CIl) transition at 809.342 GHz is roughly that of  
CO(1--0), these observations
suggest that the CO and \fCI\ emission originates in the same volume.   
This fact presents the
opportunity to examine the emission region independently of CO, in a  
pair of optically thin
lines that can be used to infer \CI\ excitation, physical conditions,  
and mass.  In EMGs, the
large redshift eliminates the burden of working at the \fCI\ rest  
frequencies, which fall in
regions where the Earth's atmosphere makes observations difficult.  
Papadopoulos, Thi \& Viti (2004) have discussed the utility of the the \fCI\ lines for the study  
of EMGs.

\fCI(\CIu $\rightarrow$ \CIll) emission has been observed in five  
ULIRGs (Gerin \& Phillips 2004; Papadopoulos \& Greve 2004), where inferred masses of molecular gas from the \fCI  
observations
assuming a relative abundance of \CI\ to H$_{2}$, $X$(\CI)$\linebreak  
=3\times 10^{-5}$, the
value inferred for M82 (Wei\ss et al. 2003), agree well with those from CO  
assuming $\alpha=0.8$,
the value usually adopted for ULIRGs.  This further supports a common  
emission region hypothesis.

If the \CI\ levels are thermally populated, then the excitation  
temperature can be calculated
from $T_{\rm ex}=38.8$K/ln(2.11/$R$\fCI), where $R$\fCI is the ratio of  
(2--1) to (1--0)
integrated line intensities (Stutski et al. 1997).  \CI\ masses can be  
calculated from

\hspace{0.5truein}$M$(\CI)$=0.911\times  
10^{-4}Q(T_{ex})e^{62.5/T_{ex}}$\Lcpu\ [\Msun],

\hspace{0.5truein}$M$(\CI)$=1.902\times  
10^{-4}Q(T_{ex})e^{23.6/T_{ex}}$\Lcpl\ [\Msun],\\
where $Q(T_{ex})=1+3e^{T_{1}/T_{ex}}+5e^{T_{2}/T_{ex}}$ is the   
partition function
(Wei\ss\ et al. 2005).

Four EMGs have been observed in \fCI\ emission: the Cloverleaf, F10214,  
SMM J14011, and PSS
J2322 (Barvainis et al. 1997; Pety et al. 2004; Wei\ss\ et al. 2003, 2005).  Only the  
Cloverleaf has been observed in
both \fCI\ lines, with an inferred excitation temperature of 30 K,  
somewhat colder than the fit
to the SED dust component of 50 K (Wei\ss\ et al. 2003). Assuming the same  
$T_{\rm ex}$ for the
Cloverleaf and F10214, and using CO data to infer the mass of \Htwo, Wei\ss\ et al. (2005) found carbon abundances for 
  all three of
$X[$\CI$]/X[\Htwo]\sim 5 \times 10^{-5}$, assuming $\alpha=0.8$, the  
ULIRG value, and ignoring
differential magnification of \fCI and CO.  The carbon abundance in PSS 
  J2322 is $3 \times
10^{-5}$ (Pety et al. 2004), close to the value for the other three  
detections.  This is an indication
of substantial enrichment in heavy elements as early as $z\sim 2.5$.  
Within the uncertainties,
there are no strong differences in the properties inferred from \fCI\   
observations between the
three QSOs and the SMG in the sample of four.

Theoretical models predict that \fCII emission in the (\CIIu  
$\rightarrow$ \CIIl) fine-structure
line at 1900.54 GHz is an important  coolant for the photo-dissociation 
  regions of molecular
clouds, more important than the emission lines of either CO, \fCI, or  
other atomic fine-structure lines.   \fCII emission has been observed in galactic  
molecular clouds, normal
galaxies, and ULIRGs.  The bulk of the extragalactic observations were  
made with the Infrared
Space Observatory (ISO) and show that ULIRGs are weaker in \fCII than  
might be expected from
a simple extrapolation from the Milky Way (for a review see Malhotra  2000).  Only upper
limits have been obtained for \fCII emission in EMGs (DJ Benford et al. manuscript submitted, van der Werf 1999).  A search for \fCII emission in SDSS J1148 
(Bolatto, Francesco \& Willott 2004) yielded an upper limit that suggests that the weakness of 
\fCII emission in ULIRGs persists to  redshifts as high as $z\sim6$.  
However, even at the
current upper limits \fCII remains the dominant coolant, roughly twice  
as important as CO and
\fCI combined (Pety et al. 2004).  This is an area where the sensitivity of  
ALMA is required for
significant progress.

\subsection{Masses, Sizes, \& Evolutionary Destiny}

Size measurements of CO emission from EMGs are constrained by the  
limited resolution and
sensitivity   of existing telescope arrays. In strongly lensed systems  
this limitation can be
overcome, and effective angular resolution of the source can be ten or  
more times greater than
the instrumental resolution of the magnified image. Derived source  
diameters then 
depend on the accuracy of  available lensing models. For most EMGs, the 
  measured CO sizes
provide only upper limits. There are a few EMGs, including two radio  
galaxies without lensing
and two SMGs,   where CO measurements indicate extended or complex CO  
morphology. There is
also indirect evidence of extended, large molecular gas disks from  
measurement of extended
nonthermal radio continuum (Chapman et al. 2004) and, by implication,  
extended FIR and CO
emission based on the radio-FIR correlation
(Carilli, Menten \& Yun 1999).  We concentrate here on  direct CO measurements of  
the size and/or separation
between the components  of the molecular gas. The CO kinematics also makes  
it possible to
estimate a dynamical mass that is independent of the gas mass  
determined from the CO line
luminosity.

The size and mass of the molecular gas disks  are important factors in  
determining the
evolutionary state  of EMGs.   Local infrared galaxies and, in  
particular, ULIRGs share many of
the properties of this high-redshift sample. They have luminosities  
greater than 10$^{12}$
\Lsun\ (Sanders \& Mirabel 1996)  and in a large sample all but one are CO  
luminous (Solomon et al. 1997)
with an average gas mass of $7\times10^9$ \Msun\ (using the conversion  
factor adopted in
Section 2.2). The molecular gas is in centrally concentrated rotating  
disks with characteristic
diameters of 0.7--2.5 kpc (Downes \& Solomon 1998) although molecular emission 
  extends out  about
twice this far.  ULIRGs result from the merger of two gas-rich spiral  
galaxies (Sanders \& Mirabel 1996)
in which the gas is driven toward the center. The large gas mass and  
presence of ample dense
molecular gas
(Gao \& Solomon 2004) lead to models where most of the FIR luminosity is  
derived from a starburst but
some of the ULIRGs are clearly composite AGN-starburst sources. Although  
the properties of
ULIRGS and EMGs overlap, many  of the EMGs are more extreme objects  
than ULIRGs with higher
IR and CO luminosities implying higher star formation rates and higher  
molecular gas mass.
This leads to  suggestions that the submillimeter population, or some portion of it,   
represents the formation of giant
($>$L*) elliptical galaxies  (Genzel et al. 2003, Greve et al. 2004a, Neri et al. 2003, Papadopoulos et al. 2000), clearly not
what is happening in ULIRGs.

\subsubsection{Summary of Molecular Gas Mass (\Htwo $+$ He)} Figure 10 
shows the gas mass
(\Htwo $+$ He) derived from the CO luminosity for the ULIRGs and EMGs  
as a function of
redshift. In cases where the magnfication has been estimated the figure 
  shows the intrinsic
mass. Otherwise a magnification of 1 is assumed. There are 11 EMGs with 
  a gas mass
essentially the same as that of local ULIRGs. As discussed in the  
previous section, most of these
(8/11) have the same or slightly higher FIR luminosities as that of ULIRGs. One 
 galaxy has a
gas mass 10 times  less than a typical ULIRG, similar to an ordinary  
spiral. This object, MS1512-cB58, is a
Lyman Break galaxy with a very large magnification 
and is clearly not a part
of the EMG population since it is not a molecular gas-rich galaxy.     
There are 21 EMGs with a
molecular gas mass significantly higher than that found in ULIRGs and higher 
  than that of any
galaxy in the local Universe. They range in gas mass from about 2.5 to  
$10\times10^{10}$ \Msun.
They include SMGs, radio  
galaxies, and molecular disks
associated with a few quasars. Some of these systems have multiple  
components and may
represent interacting or merging galaxies. A few  may have  
lensing not yet detected or
with a magnification not properly estimated.

\subsubsection{Size measurements and Dynamical Mass}

  Table 2 summarizes the observed sizes  of the CO emission regions  
excluding galaxies with
upper limits. The full range of source diameters is from 0.8 to 16 kpc  
with all but two of the
diameters falling between 1 and 5 kpc. The highly magnified CO emission 
  associated with some
quasars in Table 2 has sizes for the molecular rings or disks  
comparable to nearby ULIRGs.  The
dynamical masses listed in Table 2 have been calculated from
$\Mdyn sin^{2}i = 233.5R\Delta V^{2}$, where
$R$ is either the radius of the molecular disk or half the separation  
between components in a
merger model, measured in pc, and
$\Delta V$ is the FWHM of the CO line profile or half the separation in 
velocity of  the component CO lines in
a merger model, measured in kilometers per second.  The unknown geometry of these 
  systems precludes
more accurate estimates. Footnotes are given for those cases where this 
  calculation yields a
result differing substantially from that in the reference cited. The  
gas masses for this subset
with measured sizes are the same as in Table 1 and Figure 10.

   The largest source is the SMG J02399 with a  diameter of 16 kpc
(Genzel et al. 2003) after allowing for a magnification of 2.5 due to the
intervening cluster lens. The size is obtained from the CO data by  
fitting a
model of a rotating disk with a velocity of 420 km s$^{-1}$, a flat  
rotation curve and
a large turbulent velocity. This leads to a molecular ring with a  
maximum gas
density at $R$ = 3.2 kpc and a width of 1--1.5 kpc. The 6--8 kpc outer  
radius for
the gas also matches the  extent of the submillimeter dust continuum.  
The ring is
required to fit the double-peaked line profile. This large disk size is 
  the
total extent rather than the half power diameter which is only about 1  
kpc
larger than the peak of the ring corresponding to a half power diameter 
  of 8
kpc.  Although Genzel et al. (2003) stress the rotating molecular  
starburst ring
model with an AGN at the center of the ring,  an alternative
configuration with two galaxies orbiting each other with the AGN in  
either the
red or blueshifted CO source is possible. Indeed, the double-horned line
profile with a steep drop in the middle and the position velocity  
diagram could
easily be due to two separate galaxies, each with a much smaller  
unresolved CO
disk or ring. Thus, it is not clear if the quoted diameter  is a  
separation
between two unresolved disks or a disk size. The dynamical mass for a  
merger model is \Mdyn sin$^{2}i\sim3\times10^{11}$ \Msun.

The other SMG with a measured CO size is J14011. Ivison et al. (2001)  
found a  CO(3--2)
size of 6.6$\arcsec$ corresponding to 56 kpc in the image plane and 22  
kpc in the
source after accounting for magnification by a factor of 2.5 due to the 
  intervening
cluster Abell 1835. If real, this would have been the largest  
high-redshift galaxy found
at any wavelength.
Downes \& Solomon (2003) using the IRAM interferometer mapped both the 
  (3--2) and (7--6) lines
with high sensitivity and resolution.  They measured the peak flux to  
an accuracy of 14$\sigma $ and found an image size of 2$\arcsec\ \times 
  \leq 0.5 \arcsec$. For magnification as small as 2.5,  the intrinsic  
source diameter is reduced to less than  7 kpc. Downes \& Solomon  
(2003)
also suggested a lensing model with an intervening galaxy in addition  
to the cluster
lens. The total magnification was 25f$_{\rm v}$ where   f$_{\rm v}$ is the velocity filling factor of the CO  
emission.  This model,
with increased magnification by an intervening galaxy, has been  
questioned
(Genzel et al. 2003, Tecza et al. 2004), but Tecza et al. (2004) increased the expected  
cluster
magnification to 5.
   In Table 2 we treat the magnification of J14011 as uncertain with a  
maximum  of 25 and a
minimum of 5. This reduces the source diameter to the range of 0.7--3.5 
  kpc.  The
molecular mass is in the range of  0.4--1.7$\times10^{10}$ \Msun. The  
observed CO
spectral line is narrow with a FWHM of 190 km s$^{-1}$ (Downes \& Solomon 2003) 
  indicating a
moderate dynamical mass of about $3\times10^{10}$ \msun\ for an assumed 
  inclination
of 45$\deg$ and the larger diameter of 3.5 kpc. Unless the disk is
completely face on and/or the magnification  is much less than 5,  
the dynamical
mass is similar to  that of ULIRGs such as Mrk 231, Arp 220 , VII ZW31,  
and IR23365+36
(Downes \& Solomon 1998).

As part of a large survey of CO emission from SMGs
Greve et al. (2004a) summarized the measured  linewidth and CO  
luminosity of 11
SMGs.   They found a large median linewidth of 780 $\pm$ 330 km  
s$^{-1}$ (FWHM), 
2.5 times larger than the median width for local Universe  ULIRGs, with 
  several
examples of double-peaked profiles. The largest linewidth for a ULIRG  
in a sample of
37 galaxies is 480 km s$^{-1}$. The SMG sample also has a high median  
CO line luminosity
($3.6 \times 10^{10}$ \kkmspc)
  with a median molecular  mass of $3 \times 10^{10}$ \msun, four times 
higher  than the
ULIRG mean (Solomon et al. 1997). Although Greve et al. (2004a) concluded  
that this is
sufficient gas mass to form the stars of a giant elliptical galaxy,  it 
  seems small
unless most of the mass is already in stars and the SMGs represent a  
late stage of galaxy
formation.  The large linewidths  indicate a large but very uncertain  
dynamical mass, owing to
the absence of size measurements and unknown geometry. Assuming a  
separation (diameter ) of 3.7
kpc Greve et al. (2004a) gave a median dynamical mass
\Mdyn sin$^{2}i=1.2\times10^{11}$\ \Msun.

There are some IR
luminous interacting galaxies in the local Universe with very large  
linewidths similar to the
EMGs; one example is the LIRG Arp 118 (NGC1144) --- an unusual ring galaxy with a  
total CO linewidth  of
1100 km s$^{-1}$ and a FWHM of about 750 km s$^{-1}$. Whereas the  
linewidths of the SMG population
are similar, the SMG population is  more than an order of magnitude  
higher in luminosity.

The most impressive measurement in Table 2 is the size and structure of 
  the
CO(3--2) emission from the  $z$ = 6.4 quasar J1148+52.  Walter et al.  
(2004) mapped the CO(3--2) line with a resolution of  
0.3$\arcsec$ and  0.15$\arcsec$, the latter equivalent to about 1 kpc.  
The results show a disk with a maximum
diameter of 4.8 kpc  and a FWHM of 3.5 kpc.  The entire disk is two or three   
times as large as
a typical ULIRG.
  The core region shows two distinct sources separated by 1.7 kpc with a 
  size of roughly
0.5 kpc that account  for half of the total emission. Each of these  
regions is similar
to a nearby ULIRG in terms of mass, intrinsic brightness temperature,  
and size
(Walter et al. 2004). A detailed comparison with ULIRGs suggests that each  
of these
components may resemble the core of the molecular region in a ULIRG  
rather than the
whole disk.

Some of the high-$z$ radio galaxies show kinematic structure indicating
the presence
of two merging galaxies. In 4C41.17 (De Breuck et al. 2004) the two CO  
components are
separated by 1.8$\arcsec$ or 13 kpc with a velocity difference of 500  
km s$^{-1}$. Each
component has a molecular mass of about $3\times10^{10}$ \Msun. This 
system
appears to be a major merger  in progress between two gas-rich galaxies 
  rather than
one extended very massive disk.  Each component remains unresolved. The
dynamical mass of the system is \Mdyn sin$^{2}i = 6 \times 10^{11}$  
\Msun.

4C60.07 also shows possible evidence of an ongoing merger between two  
galaxies although the
angular separation between the components is not well determined.  
Papadopoulos et al. (2000)
imaged the CO(4--3) line and found an extent or separation of  
7$\arcsec$ or 51 kpc, but the resolution
of the measurements was only 9$\arcsec \times  5.5\arcsec$. Higher  
resolution measurements in the
CO(1--0) line (Greve et al. 2004) show a separation of 4$\arcsec$ or about  
28 kpc in the images tapered
to 60 k$\lambda$; the higher resolution images tapered to 200 k$\lambda$ 
  show a smaller angular
separation of only about 1$\arcsec$. Using the larger separation they  
calculate a total dynamical
mass  between 0.2 and $0.8 \times 10^{12}$ \msun\ comparable to the 
mass of  a giant elliptical galaxy.

In Table 2 we list two size ranges for ULIRG molecular disks in the local Universe, including the half-power diameter and the total diameter for CO emission.  The measured diameters of EMGs fit within the range measured for ULIRGs with one noticeable exception.  The total gas mass of EMGs covers a wide range.  About half of the EMGs have a total gas mass above that found for any ULIRG and, thus, represent the largest reservoirs of star-forming molecular gas in the Universe.

\subsubsection{Are EMGs Massive Galaxies in Formation?}

  A critical question is whether the EMGs or some fraction of the EMGs  
represent the formation
of massive galaxies in the early Universe. The star formation rates  
derived from the FIR
luminosity range from about 300 to 5000 \Msun year$^{-1}$ (see Figure 8 
). At the lower  end, these
star formation rates are similar to local ULIRGs and represent 
starbursts in  centrally concentrated
disks sometimes but not always associated with AGNs. These events may  
form a central bulge
but not a giant elliptical galaxy. At the higher end, it would take  
10$^{8}$ years to produce a
stellar mass of 3--5$\times10^{11}$ \Msun\ typical of the stellar mass 
of  a giant elliptical galaxy.
This is a reasonable time scale. However, the available molecular gas  
supply, about 3--6
$\times 10^{10}$
\Msun, falls short by a factor of 5--10. The gas lifetime is too  
short. The remaining 80--90\% of the mass would already have to be in the stellar component of  
the EMGs or added later
by subsequent mergers in order to account for the formation of a giant  
elliptical galaxy. Accurate
measurements of the dynamical mass and size scale of EMGs are needed to 
  provide convincing
evidence for  EMG masses similar to modern elliptical galaxies.  Table  
2 shows five EMGs with the
approximate dynamical mass in the right range, but the size  
measurements are only marginally
significant in most cases. The large linewidths of the SMG population
(Greve et al. 2004a) are a good indication that the total mass of some of  
these early galaxies is
large but most of these have unknown morphology and do not have size  
measurements. CO
images with substantially higher resolution and sensitivity are  
required.

The EMG population clearly represents a major stage in galaxy formation.
The high star
formation rates, high total molecular mass, and, in some cases, high  
mass of dense molecular
gas all point to huge starbursts,  much  greater than   observed in  
individual  optical-UV
starbursts.

\subsubsection{Comparison with Lyman Break Galaxies}

  The distribution of star formation rates from LBGs  
obtained directly from
the UV flux shows a peak at about 20  \Msun\ year$^{-1}$ falling off  
rapidly for higher star formation rates
(Giavalisco, 2002). Correction for extinction involving dust  
scattering models and
stellar population synthesis shifts the peak to about 100
\Msun\ year$^{-1}$ with a broad distribution and a tail extending up to  
about 700  \Msun\ year$^{-1}$.
In the most extreme cases the UV radiation  captures much less than  
10\% of the total
luminosity with the rest shifted into  the infrared. Tests of this  
extinction correction technique
show that it fails completely for local ULIRGs  (for example, 
Giavalisco 2002)  and would also fail for
EMGs. The range of star formation rates for EMGs begins at the  higher  
end of the extinction-corrected UV values for LBGs and extends upward by  an  
order of
magnitude. Giavalisco (2002) suggests that the star formation observed  
in LBGs  could lead after 1 Gyr to an L* galaxy.  But there is no  
evidence for the
presence of sufficient interstellar gas in LBGs to  
build up an L* galaxy. The
one LBG found with CO emission, MS1512-cB58 (see Table 
2),  contains only a few
$\times\ 10^{8}$ \Msun\ of molecular gas, 30 times less than the mean 
of the EMG  sample (this EMG
appears as a low outlier in Figure 6), and more than two orders of  
magnitude below the mass of an
L* galaxy.

The total contribution to early Universe star formation from 
SMGs compared with
LBGs depends on an understanding of the origin of the  
FIR
background, a topic addressed elsewhere (Puget \& Lagache, this volume).

\section{ OBSERVATIONAL PROSPECTS}

Observations of the molecular gas discussed here are critical for  
understanding early Universe
galaxy formation. The morphology, kinematics, and gas density estimates 
  provided by better
measurements of CO and other molecular lines will lead to a detailed  
understanding of the
processes and mechanisms involved in assembling galaxies and forming  
stars in the early
Universe.

The present suite of telescopes available for the detection of EMGs has 
  produced a sample of
36, which is expected to grow, particularly for SMGs, within the limits of the  
observing time allocated
for high-$z$ CO emission searches.  A doubling of the sample is not  
unreasonable to expect in
the next five years.  But this falls far short of the sample sizes  
needed for true statistical
studies of EMG properties.  The current sample is especially deficient  
at redshifts $z>3$, where
the potential of the EMGs for the study of galaxy formation is most  
important.  There is only
one EMG that probes the era of re-ionization.

Besides their limitations for the detection of more EMGs, the ability  
of the present telescopes
to study these objects in detail is severely limited in sensitivity and 
  angular resolution.  Only
the strongest sources, observed at high frequencies, possibly through  
gravitational lenses, and
with long integration times, offer clues regarding the structure of EMGs. To understand EMGs,  images that resolve and map the molecular  
line emitting region are critical.

ALMA is the only observing facility planned for operation within the  
next decade that combines
the sensitivity, angular resolution, flexibility of observing modes,  
and site conditions required
for such imaging.  ALMA will be the premier telescope for the study of  
EMGs. Its 64
12m-diameter antennas provide the collecting area needed for high  
sensitivity.  The ability to
reconfigure the array allows one to select angular resolution for any  
observing frequency. The
angular resolution at a frequency of 350 GHz is 1$\arcsec$ in the  
compact configuration, as high as
0.014$\arcsec$ using baselines up to the maximum of 14 km, and scaling  
inversely with
frequency.  The correlator can process up to 16 GHz of bandwidth from  
each antenna, in four
separately tunable 2-GHz-wide signals in each of the two polarizations.   
The receiver noise will be
three times the quantum limit ($T_{\rm rx}\approx3h\nu/k$) for all but the highest frequency receiver bands.  A compact 
array of 12 7m-diameter antennas, plus four 12m diameter antennas 
for calibration purposes,  bolsters sensitivity on spatial frequencies 
between that of a single 12m antenna and the shortest baseline (15m) in the 
large array.  The site is  comparable in quality to
the South Pole for millimeter/submillimeter observing, and superbly  
located for studying the southern sky and much of the northern sky. For further information on  
ALMA, the reader is
referred to http://www.alma.nrao.edu/ and http://www.eso.org/projects/alma.

Guilloteau (2001) and Blain (2001) have reviewed ALMA's capability to observe high-$z$ spectral line and continuum emission, respectively.  As an  
illustration of ALMA's
power for detailed studies of EMGs, consider the SMG J23099, where the  
CO source has been
modeled (Genzel et al. 2003) as a rotating disk of diameter 16 kpc  
(5$\arcsec$).  When used in a 6-km-maximum baseline configuration (resolution 0.5$\arcsec$)   
with an 8-hour integraton, ALMA will
yield an image with velocity resolution of 100 km s$^{-1}$ and rms  
noise of 0.4 mJy
(5$\sigma$).  This is 10\% of the unresolved flux density of the source,  
enough to  check the
validity of the model.  Because this observation can be done with only  
one of the tunable 2-GHz
inputs to the correlator, simultaneous observations of, say, CS(7--6),  
HCN(4--3), and up to 29
other lines within the instantaneous bandpass of the receiver could be  
made.  Although these lines
may not be detected in a single 0.5$\arcsec$ beam, the u-v data, fully  
sampled to 6 km, could
be smoothed to 1$\arcsec$ resolution, thereby yielding a 5$\sigma$ sensitivity  
of 0.1 mJy.

For simple detection of EMGs in CO emission,  the (6--5) transition, for example,   
at a redshift of $z$=2
with a peak line intensity of 1 mJy beam$^{-1}$ (or any spectral line  
in the bandpass with this
peak line strength), would be seen by ALMA at the 10$\sigma$\ level  
with velocity resolution
of 50 km s$^{-1}$ in a typical 4-h observing session. The continuum 
  emission observed in
this same session at 230 GHz would reach a 5$\sigma$\ sensitivity of 33 
  $\mu$Jy
beam$^{-1}$.  The continuum emission from Arp 220 moved to a redshift  
of $z$=2 could be
detected at the 5$\sigma$\ level in less than 30 min of observing  
time.  Because of the
``negative K-correction,'' this statement is true for Arp 220 at any  
redshift up to $z\sim20$.

Given the sensitivity of ALMA, with seven times the collecting area of  
the IRAM interferometer
and a superior site, it is clear that the study of EMGs will be  
transformed from one of imaging
CO emission to one of imaging emission from a variety of interstellar  
molecules.  The importance to gas density studies of
HCN, [\CI], and [\CII] have been discussed above.  Carbon monosulfide may be an even
better tracer of dense, star-forming gas than is HCN (Shirley et al. 2003, but 
  its weaker lines remain
beyond the reach of present telescopes. Formaldehyde  is  
another molecule that
traces dense gas, potentially accessible to ALMA observers of EMGs.   
Searches should
be made with ALMA for the isotopomers of CO.  The ALMA correlator can observe many lines simultaneously, making it very powerful for astrochemical studies.

The potential for ALMA to reveal the process of galaxy formation and  
evolution in the early
Universe can be summarized by noting that observing CO emission in the 
$z$=6.4 quasar SDSS J1148  tests limits
of present instruments. ALMA will be able to observe CO in a galaxy at 
this redshift having the CO luminosity of a large, normal spiral such as 
M51 or NGC 891, making it possible to probe the era of re-ionization 
with a much larger population.  Readers who wish to design their own 
ALMA observing  programs can find a
sensitivity calculator at  
http://www.eso.org/projects/alma/science/bin/sensitivity.html.

Other facilities will also play a significant role in the study of  
EMGs.  An upgraded IRAM
interferometer, the Combined Array for Research in Millimeter-wave Astronomy (CARMA), the Submillimeter Array (SMA), and the Extended VLA (EVLA) will add increased  
sensitivity and/or
bandwidth to present capability.  For objects with redshift $z\geq2$,  
CO emission from low-J levels falls in the centimeter wavelength observing bands of the EVLA. 
   The EVLA will be
particularly suitable for observing HCN in lower-J transitions.   
Receiver systems working to
wavelengths as short as 0.7 cm combined with a powerful wide-band  
correlator will make the
EVLA a powerful telescope  for EMG observing in the Northern  
Hemisphere. Large single dishes
such as the Green Bank Telescope (GBT) are also proving useful for EMG  
study, as the detection of
HCN(1--0) emission in F10214 (Vanden Bout, Solomon \& Maddalena 2004) has demonstrated.   
The GBT will be
primarily useful for measuring CO(1--0) luminosity, detecting new EMGs  
in that line, and doing
continuum surveys with 3 mm wavelength bolometer cameras.  Upon  
completion, the 50-m
diameter Large Millimeter Telescope (LMT) will be the most powerful 
single-aperture telescope for the study of
EMGs. Its very substantial collecting area will make it a  
telescope of choice for blind surveys.

The next decade will see explosive growth in the number of known EMGs,  
the findings concerning their properties, and
most important, in knowledge of their structure and evolution.  The  
ability of ALMA to image
the kinematics of the molecular star-forming gas in galaxies from the  
era of recombination
to the present will be invaluable to our understanding of the evolution 
  of galaxies and the Universe.

\scshape ACKNOWLEDGEMENTS

\rm We gratefully acknowledge the assistance of J. W. Barrett in the preparation of the figures.  PVB is grateful for the  hospitality of the Institut d'Astrophysique, Paris, and the Department of Astronomy, University of Texas, Austin, during the writing of this review.

\begin{figure}
\epsfig{file=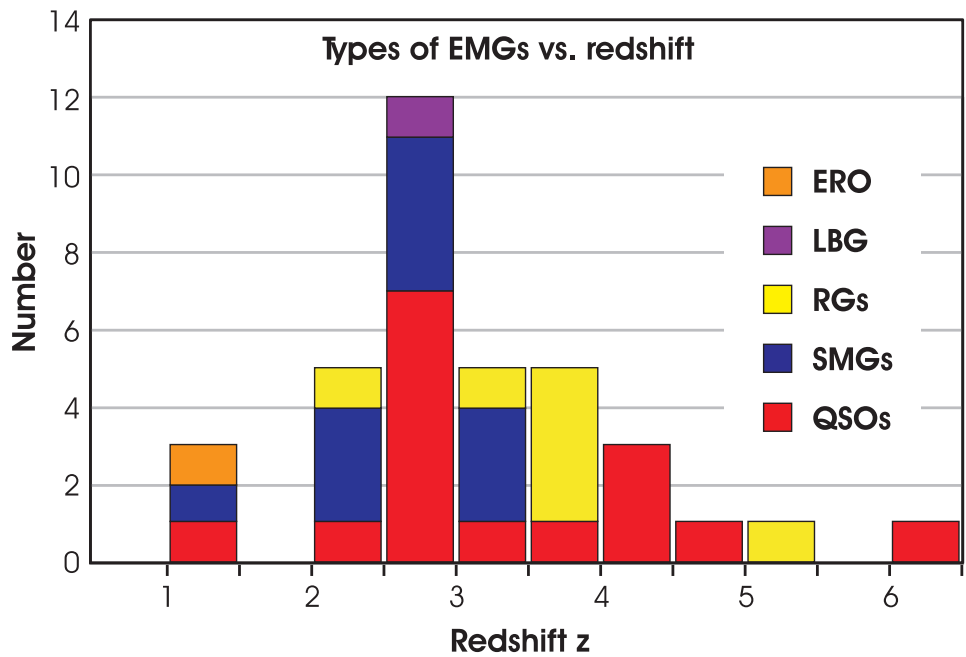,width=5truein}
\caption{Distribution in redshift of the 36 known EMGs: 16 quasi-stellar objects (QSOs), 11 submillimeter galaxies (SMGs), 7 radio galaxies (RGs), one Lyman Break galaxy (LBG), and one extremely red object (ERO).  Despite the large selection effects of the flux-limited sample, the distribution broadly reflects the current understanding of when most of the star formation in the Universe occured.}
\label{zhisto}
\end{figure}

\begin{figure}
\epsfig{file=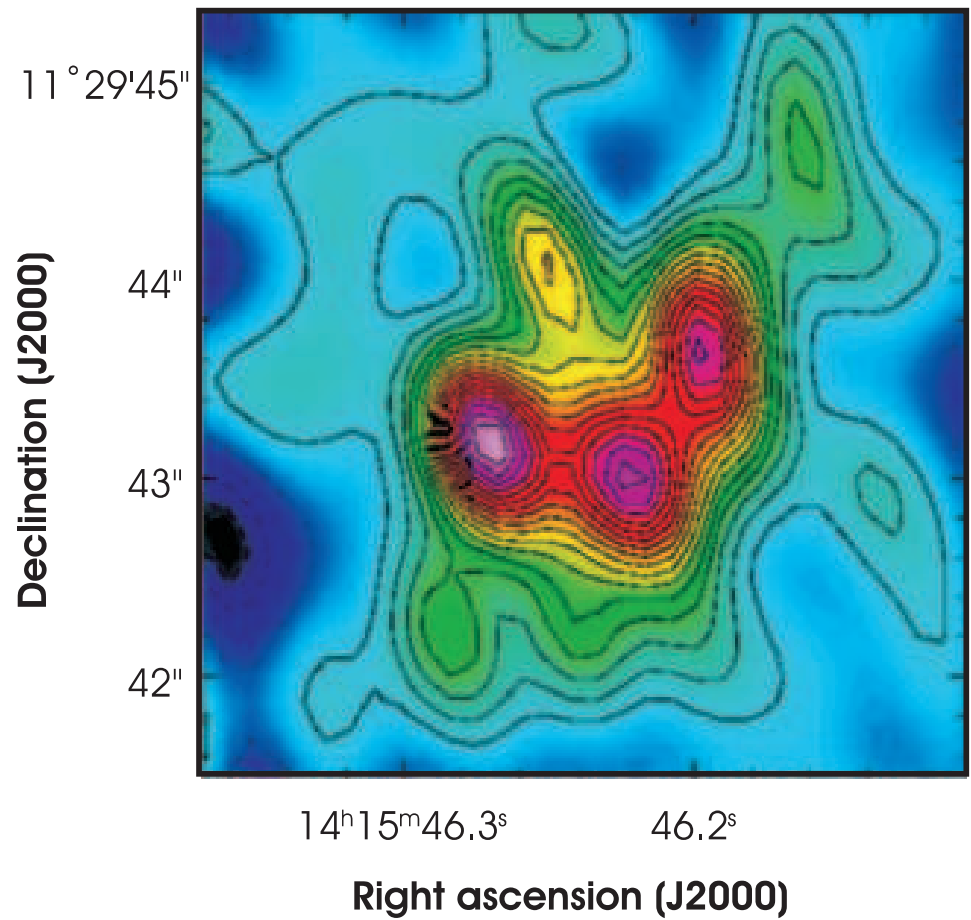,width=5truein}
\caption{Image of the Cloverleaf in CO(7--6) emission taken with the 
IRAM interferometer (constructed by Venturini \& Solomon 2004 from the data of Alloin et al. 1997).  The high observing frequency of 226 GHz provides 
the angular resolution (0.5$\arcsec$) needed to construct a 
gravitational lens model based on CO data.}
\label{clvrleaf7-6}
\end{figure}

\begin{figure}
\epsfig{file=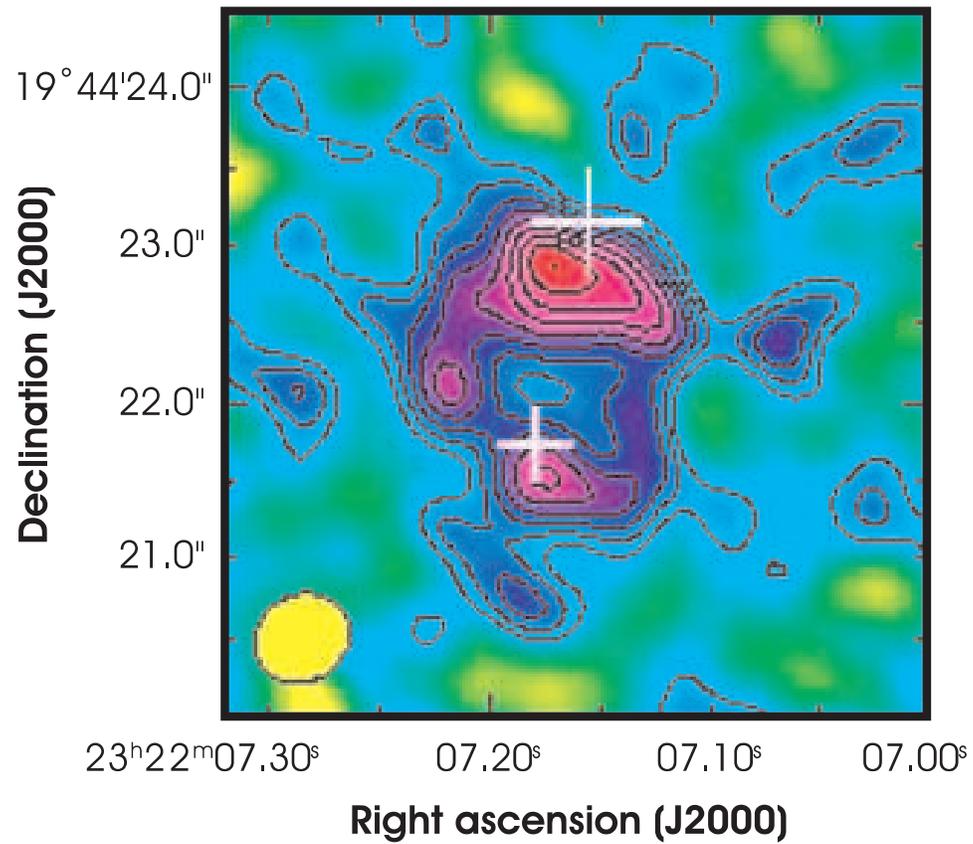,width=5truein}
\caption{The Einstein Ring in PSS2322, observed in CO(2--1) emission 
using the VLA at a resolution of 0.6$\arcsec$ (Carilli et al. 2003).}
\label{PSSeinsteinring}
\end{figure}

\begin{figure}
\epsfig{file=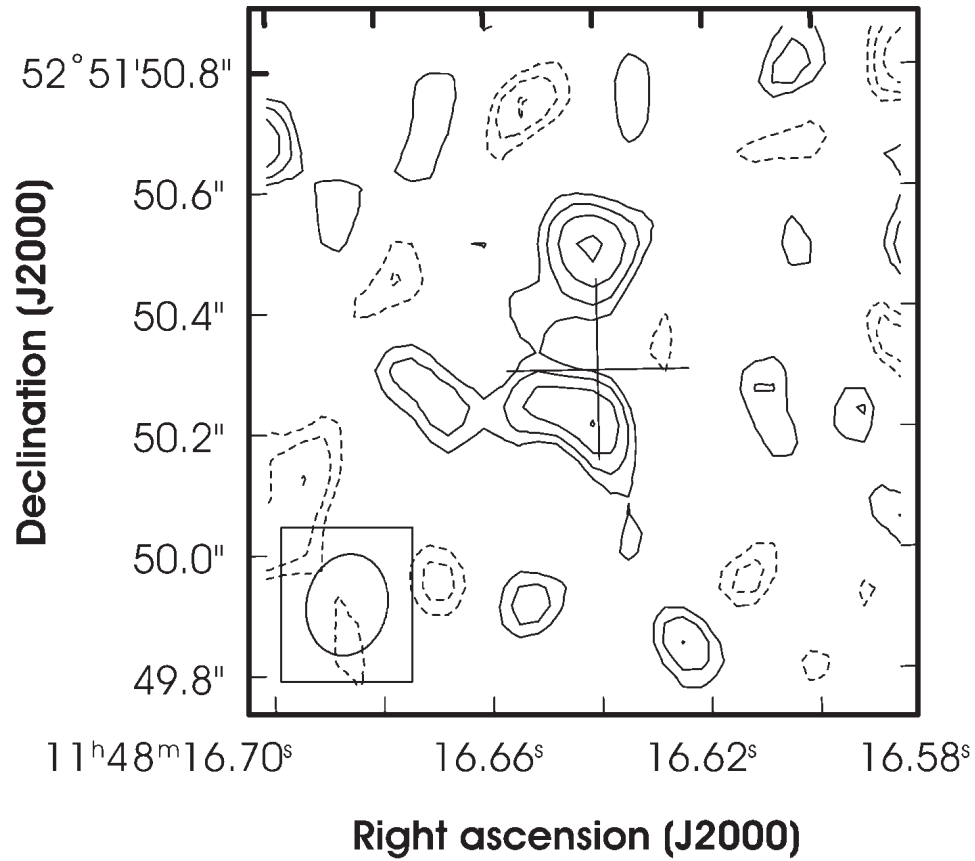,width=5truein}
\caption{SDSS J1148, a quasar at $z=6.4$ imaged in CO(3--2) emission 
using the VLA at a resolution of $0.17\arcsec\times0.13\arcsec$ (Walter et al. 2004).  This 
system is a possible merger of two components that resemble the ULIRGs 
of the more local Universe.  The presence of CO in this system is 
evidence for substantial enrichment in heavy metals $\sim$850 million 
years after the Big Bang.}
\label{1148VLA}
\end{figure}

\begin{figure}
\epsfig{file=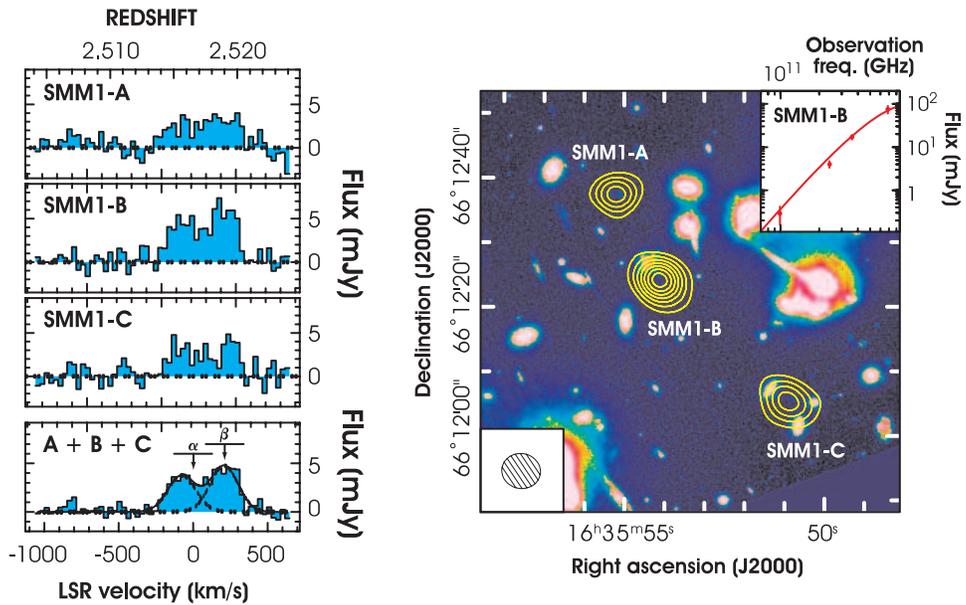,width=5truein}
\caption{The lower panel shows SMM J16399 in CO(3--2) emission that has been triply imaged by a gravitational lens (Kneib et al. 2004a).  The total magnification is $\mu=45$, making possible this observation of CO in a somewhat less luminous SMG.  the CO contours are superimposed on an HST image of Abell 2218, and show good registration with their optical counterparts.  The synthesized CO beam ($\sim6\arcsec$) is shown in the lower left corner.  The SED in the range 450--3000$\mu$m is shown in the upper right corner (Kneib et al. 2004b).  The upper panel shows the CO spectra from each image together with the combined spectrum.  The redshifts deduced from HST imaging and H$\alpha$ spectroscopy, shown as $\alpha$ and $\beta$, are in close agreement with those of the CO emission peaks.}
\label{tripleSMG}
\end{figure}

\begin{figure}
\epsfig{file=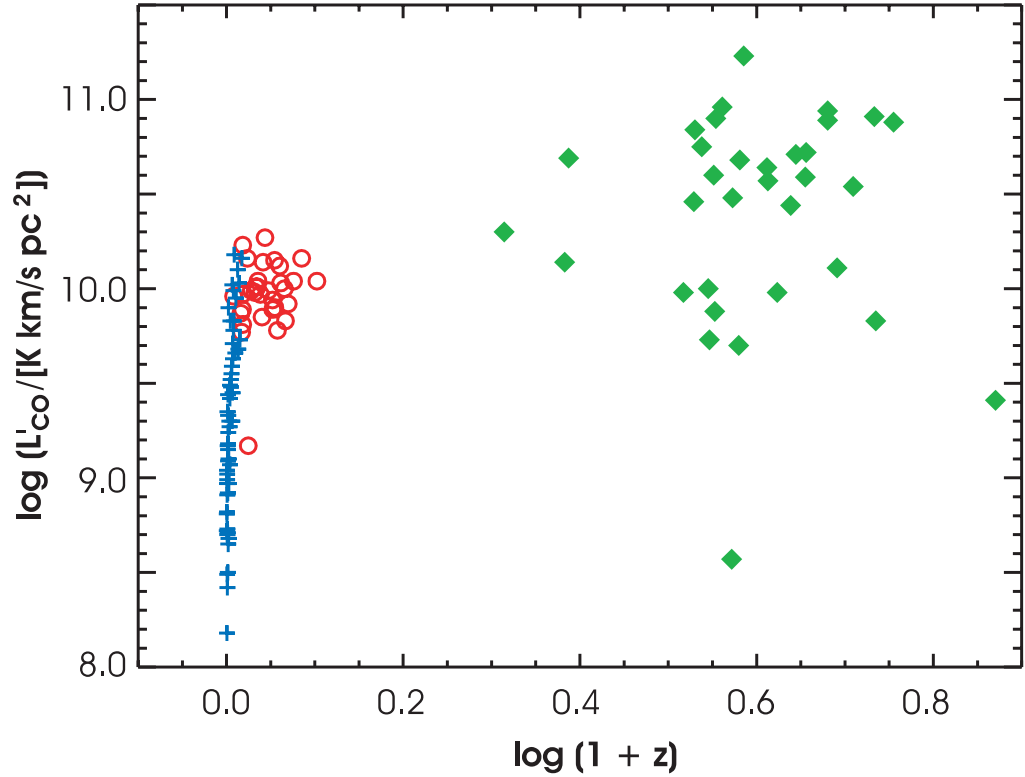,width=5.3truein}

\caption{CO Luminosity: log$\Lcop$ versus log$(1+z)$ for local galaxies 
with \Lfir $<10^{11.8}$
({\it blue crosses}),  ULIRGs ({\it red circles}), and EMGs ({\it green diamonds}).  Although
the EMGs are a flux-limited sample, the large scatter among the EMGs
shows that they are much more diverse in CO luminosity
and three times stronger in the mean compared with ULIRGs. The mean for
ULIRGs and EMGs is 1$\times$10$^{10}$ and  3$\times10^{10}$ \Kkmspc, respectively. All EMG
luminosities with known lensing are corrected for magnification.}
\label{Lprimez}
\end{figure}

\begin{figure}
\epsfig{file=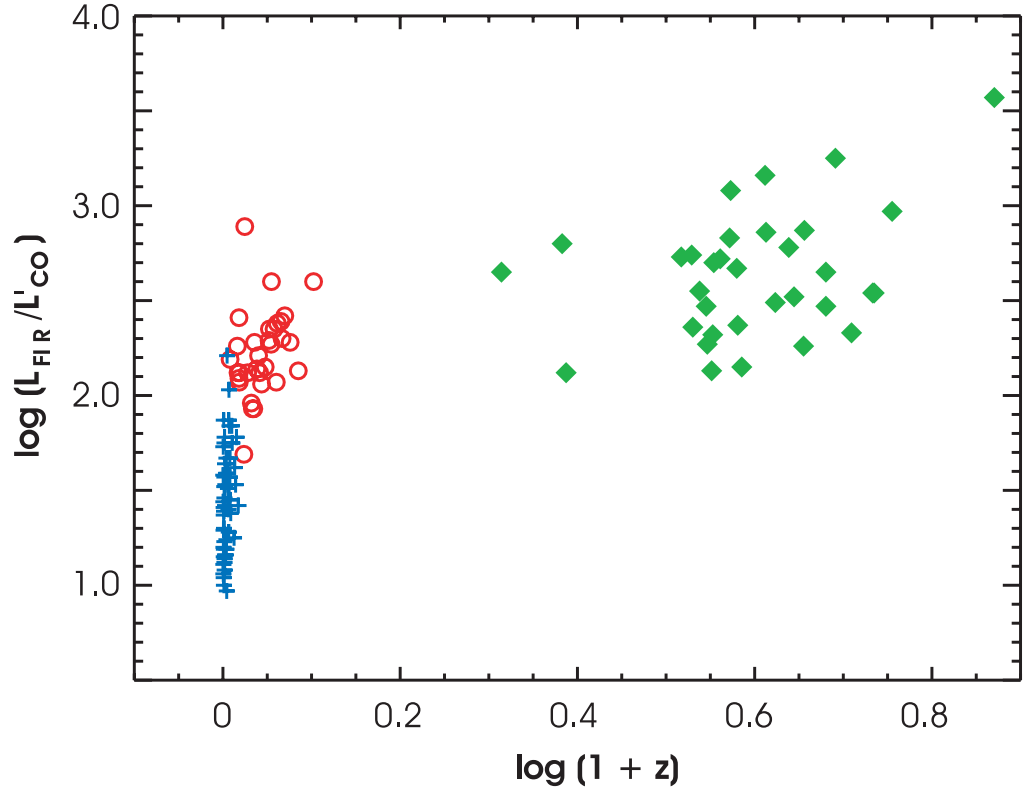,width=5.3truein}

\caption{Star Formation Efficiency: log$(\Lfir/\Lcop)$, an indicator of star formation efficiency, versus log$(1+z)$  
 for normal spirals including luminous but not ultraluminous
galaxies ({\it blue crosses}),  ULIRGs ({\it red circles}), and EMGs ({\it green diamonds}). ULIRGs 
and
EMGs both have much higher star formation efficiency (SFE) than lower 
luminosity
galaxies.   EMGs have only a factor of two higher SFE on average than
the EMGs, but there is substantial overlap even though the average FIR
luminosity and star formation rate is 10 times higher for EMGs than 
ULIRGs.}
\label{sfrz}
\end{figure}

\begin{figure}
\epsfig{file=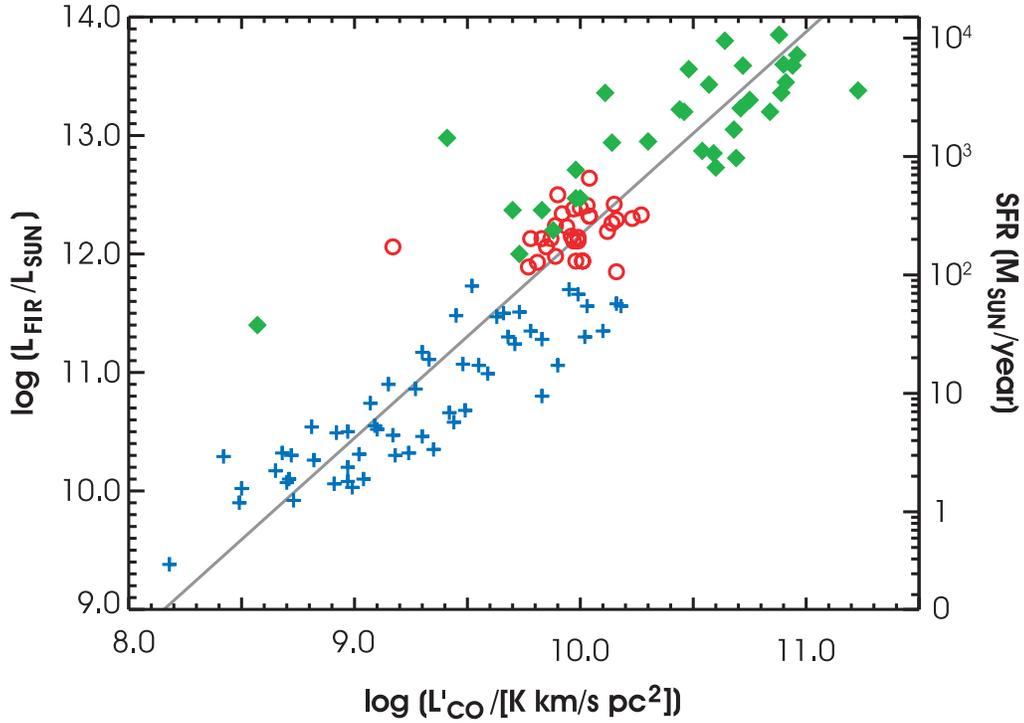,width=5.3truein}

\caption{CO as a tracer of star formation rate: log$\Lfir$ and SFR versus 
log$\Lcop$ for normal spirals
({\it blue crosses}),  ULIRGs ({\it red circles}), and EMGs ({\it green diamonds}).   The solid line,
a fit to all the points has a steep slope, log$\Lfir$ = 1.7 log$\lcop -5.0
$, showing that total molecular mass indicated by CO luminosity is not 
a linear tracer of the star formation rate, indicated by FIR 
luminosity,
  when ULIRGs and EMGs are included. Excluding ULIRGS and EMGs the
slope is 1.1.  Unlike CO, HCN luminosity is a linear tracer of FIR 
luminosity and the star formation
rate (see Section 3.2.1).}
\label{firco}
\end{figure}

\begin{figure}
\epsfig{file=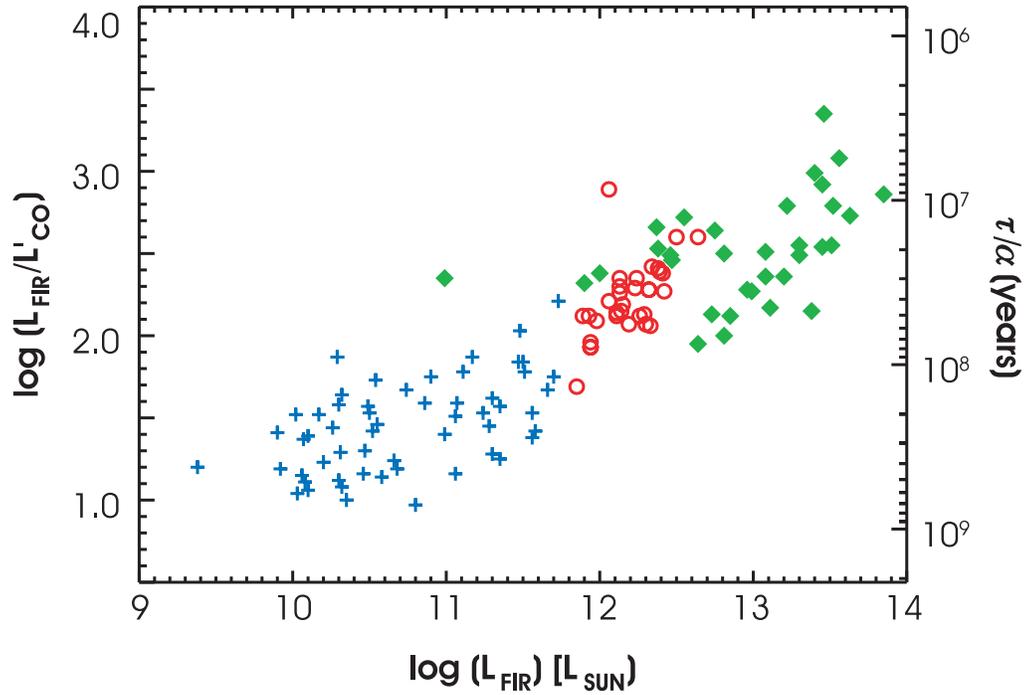,width=5.3truein}
\caption{Star formation lifetime: star formation lifetime $\tau$ due to 
gas depletion versus log$\Lfir$
for normal spirals ({\it blue crosses}),  ULIRGs ({\it red circles}), and EMGs 
({\it green diamonds}).   $\alpha$ is the CO line
luminosity to
\Htwo\ mass conversion factor (see Section 2.2), which is about 0.8 for 
ULIRGS, probably 0.8 for EMGs
and 4.6 for normal spirals. The EMG star formation lifetime is  between 
$10^7 $ and $10^8$ years (see the
text). The star formation rate can be taken as 1.5$\times10^{-10}$ \Lfir 
[\Msun\ year$^{-1}$].}
\label{tfir}
\end{figure}

\begin{figure}
\epsfig{file=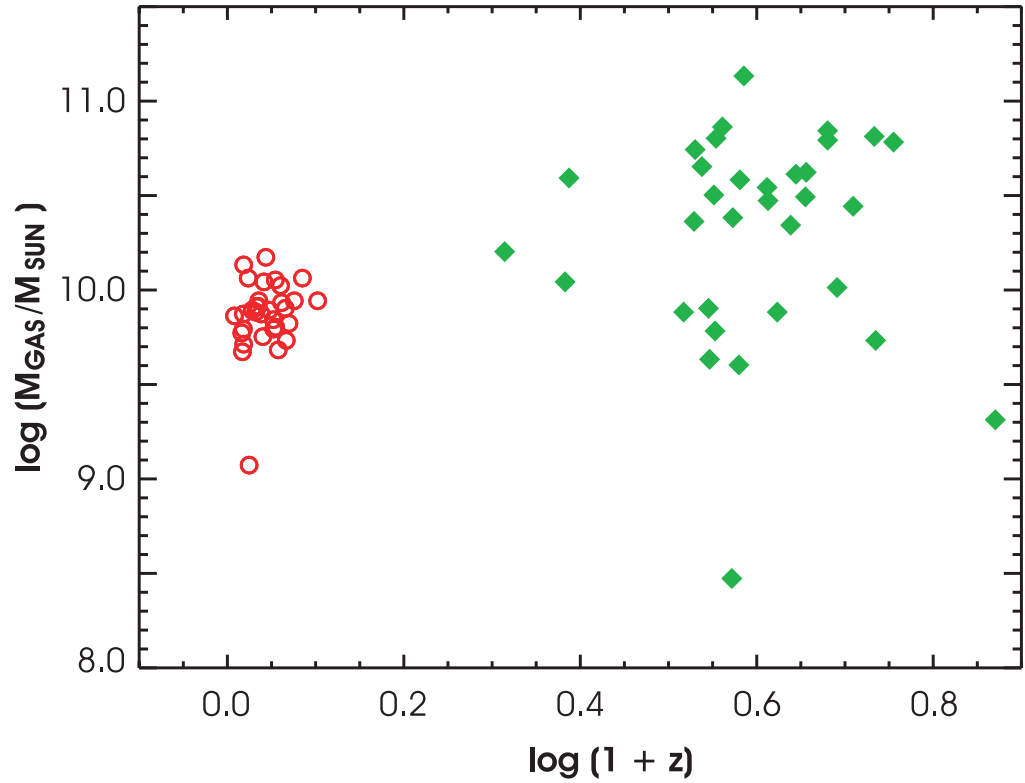,width=5.3truein}

\caption{Molecular gas mass in ULIRGs and EMGs: log$M_{\rm gas}$ versus 
log$(1+z)$
for ULIRGs ({\it red circles}) and EMGs ({\it green diamonds}). M$_{\rm gas}$ includes \Htwo\ 
and
He.   The  EMGs are more massive on average than the ULIRGs, although 
there
is considerable scatter among individual EMGs and substantial overlap 
with
ULIRGs (see the text).}
\label{massz}
\end{figure}

\cleardoublepage

\begin{landscape}
\begin{table}
\footnotesize
\def~{\hphantom{0}}
\caption{EMG Properties: line \& FIR luminosities, gas \& dust masses, 
star formation rate}
\smallskip
\begin{tabular}{lcr@{ }lr@{$\pm$}lccccllll}\hline\hline
\multicolumn{1}{c}{EMG} & Redshift & \multicolumn{2}{c}{Transition} & 
\multicolumn{2}{c}{$L^{\prime}(\rm app.)$} & $L_{\rm FIR}(\rm app.)$ & Lens & 
$L^{\prime}(\rm int.)$ & $L_{\rm FIR}(\rm int.)$ & \multicolumn{1}{c}{$M_{\rm gas}$} & 
\multicolumn{1}{c}{$M_{\rm dust}$} & \multicolumn{1}{c}{SFR} & \multicolumn{1}{c}{$\tau_{\rm SF}$}\\
& \multicolumn{1}{c}{$z$} & & & \multicolumn{2}{c}{(10$^{10}\ 
$L$^{\prime}_*)^a$} & (10$^{12}\ $L$_{\odot}$) & Mag. & (10$^{10}\ 
$L$^{\prime}_*)^a$ & (10$^{12}\ $L$_{\odot}$) & (10$^{10}\ 
$M$_{\odot}$) & (10$^{8}\ $M$_{\odot}$) & (M$_{\odot}\ $y$^{-1}$)& ($10^6$ y) \\ 
\hline
SMM J02396 & 1.062 & CO & 2--1 & 5.1 & 0.5 & 16.3 & 2.5 & 2.0 & 6.5 & 
1.6 & & 975 & 16\\
Q0957+561 & 1.414 & CO & 2--1 & 0.9 & 0.5 & 14 & 1.6 & 0.6 &6 & 0.4 & 
2.5 & 900 & 4\\
HR10  & 1.439 & CO & 1--0 & 6.5 & 1.1 & 6.5 & ? & --- & --- & 
5.2$\mu^{-1}$ & 6.8$\mu^{-1}$ & &\\
IRAS F10214 & 2.286 & CO & 3--2 & 11.3 & 1.7 & 60 & 17 & 0.7 & 3.6 & 
0.6 &&540 & 11\\
SMM J16371 & 2.380 & CO & 3--2 & 3.0 & 0.6 & --- & ? & --- & --- & 
2.4$\mu^{-1}$ &&&\\
SMM J16368 & 2.385 & CO & 3--2 & 6.9 & 0.6 & 16 & ? & --- & --- & 
5.5$\mu^{-1}$ &&&\\
53W002 & 2.393 & CO & 3--2 & 3.6 & 0.4 & --- & 1 & 3.6 & --- & 2.9 &&&\\
SMM J16366 & 2.450 & CO & 3--2 & 5.6 & 0.9 & 20 & ? & --- & --- & 
4.5$\mu^{-1}$ &&&\\
SMM J04431 & 2.509 & CO & 3--2 & 4.5 & 0.6 & 13 & 4.4 & 1.0 & 3 & 0.8 
&&450 & 18\\
SMM J16359 & 2.517 & CO & 3--2 & 18.9 & 0.8 & 45 & 45 & 0.4 & 1 & 0.3 & 
2 &150 &20\\
Cloverleaf & 2.558 & CO & 3--2 & 44 & 1 & 59 & 11 & 4.0 & 5.4 & 3.2 & 
1.5 &810 & 40\\
SMM J14011 & 2.565 & CO & 3--2 & 9.4 & 1.0 & 20 & 5--25 & 0.4--1.9 & 
0.8--4.0 & 0.3--1.5 & 0.13--0.65 &120--600 & 25\\
VCV J1409 & 2.583 & CO & 3--2 & 7.9 & 0.7 & 35 & ? & --- & --- & 
6.3$\mu^{-1}$ & 38$\mu^{-1}$ &&\\
LBQS 0018 & 2.620 & CO & 3--2 & 5.4 & 0.9 & 33 & ? & --- && 
4.3$\mu^{-1}$&&&\\
MG0414 & 2.639 & CO & 3--2  & \multicolumn{2}{c}{9.2} & 32 & ? & --- && 
7.4$\mu^{-1}$ &&&\\
MS1512-cB58 & 2.727 & CO & 3--2 & 1.4 & 0.3 & 3.1 & 32 & 0.043 & 0.1 & 
0.03 &&15 & 20\\
LBQS 1230 & 2.741 & CO & 3--2 & 3.0 & 1.0 & 36 & ? & --- & --- & 
2.4$\mu^{-1}$ & 11$\mu^{-1}$ &&\\
RX J0911.4 & 2.796 & CO & 3--2 & 11.3 & 4.3 & 51 & 22 & 0.52 & 2.3 & 
0.4 &&345&12\\
SMM J02399 & 2.808 & CO & 3--2 & 12.2 & 1.6 & 11 & 2.5 & 4.9 & 4.4 & 
3.9 & 6--8 &660&59\\
SMM J04135 & 2.846 & CO & 3--2 & 22 & 5 & 31 & 1.3 & 17 & 24 & 13.0 & 
18 &3600&36\\
B3 J2330 & 3.092 & CO & 4--3 & 3.4 & 0.8 & 28 & 1 & 3.4 & 28 & 2.7 
&&1950&14\\
SMM J22174 & 3.099 & CO & 3--2 & 3.7 & 0.9 & 12 & ? & --- && 3.0 
&&1800&17\\
MG0751 & 3.200 & CO & 4--3 & 16 & 1 & 49 & 17 & 1.0 & 2.9 & 0.8 &&435&18\\
SMM J09431 & 3.346 & CO & 4--3 & 3.2 & 0.3 & 20 & 1.2 & 2.7 & 12 & 2.2 
&&1800&12\\
\end{tabular}
\end{table}

\begin{table}
\footnotesize
\def~{\hphantom{0}}
\smallskip
\begin{tabular}{lcr@{ }lr@{$\pm$}lccccllll}\hline\hline
\multicolumn{1}{c}{EMG} & Redshift & \multicolumn{2}{c}{Transition} & 
\multicolumn{2}{c}{$L^{\prime}(\rm app.)$} & $L_{\rm FIR}(\rm app.)$ & Lens & 
$L^{\prime}(\rm int.)$ & $L_{\rm FIR}(\rm int.)$ & \multicolumn{1}{c}{$M_{\rm gas}$} & 
\multicolumn{1}{c}{$M_{\rm dust}$} & \multicolumn{1}{c}{SFR} & \multicolumn{1}{c}{$\tau_{\rm SF}$}\\
& \multicolumn{1}{c}{$z$} & & & \multicolumn{2}{c}{(10$^{10}\ 
$L$^{\prime}_*)^a$} & (10$^{12}\ $L$_{\odot}$) & Mag. & (10$^{10}\ 
$L$^{\prime}_*)^a$ & (10$^{12}\ $L$_{\odot}$) & (10$^{10}\ 
$M$_{\odot}$) & (10$^{8}\ $M$_{\odot}$) & (M$_{\odot}\ $y$^{-1}$)& ($10^6$ y) \\
\hline
SMM J13120 & 3.408 & CO & 4--3 & 5.2 & 0.9 & 12 & ? & --- & & 
4.2$\mu^{-1}$&&&\\
TN J0121 & 3.520 & CO & 4--3 & 5.4 & 1.0 & 7 & 1 & 5.4 & 7 & 4.3 
&&1050&41\\
6C1908 & 3.532 & CO & 4--3 & 5.2 & 1.0 & 9.8 & 1 & 5.2 & 9.8 & 4.2 
&&1470&29\\
4C60.07b & 3.791 & CO & 1--0 & 8.7 & 1.7 & 13 & 1 & 8.7 & 13 & 7.0 
&&1950&36\\
4C60.07r & 3.791 & CO & 1--0 & 5.2 & 0.6 & & 1 & 5.2 & & 4.2 &&&\\
4C60.07b & & & 4--3 & 6.0 & 0.9 & & 1 & 6.0 &&&&\\
4C60.07r & & & 4--3 & 3.0 & 0.2 & & 1 & 3.0 &&&&\\
4C41.17R & 3.796 & CO & 4--3 & 4.3 & 0.5 & 20 & 1 & 4.3 & 20 & 3.4 & 
4.6 &3000&11\\
4C41.17B & 3.796 & CO & 4--3 & 2.2 & 0.5 & & 1 & 2.2 & & 1.8 &&&\\
APM 08279 & 3.911 & CO & 1--0 & 9.1 & 2.7 & 200 & 7 & 1.3 & 29 & 1.0 & 
5.8 &4350&2\\
PSS J2322 & 4.119 & CO & 1--0 & 12 & 5 & 23 & 2.5 & 5.0 & 9.3 & 4.0 
&&1800&22\\
BRI 1335N & 4.407 & CO & 2--1  & 3.3 & 1.1 & & ? & --- & --- 
&2.6$\mu^{-1}$ &&&\\
BRI 1335S & 4.407 & CO & 2--1 & 4.8 & 1.1 & & ? & --- & --- & 
3.8$\mu^{-1}$ &&&\\
BRI 1335 & & CO & 5--4 &  8.2 & 0.9 & 28 & ? & --- & --- & & 
17$\mu^{-1}$&&\\
BRI 0952 & 4.434 & CO & 5--4 & 2.8 & 0.3 & 9.6 & 4 & 0.7 & 2.4 & 0.5 & 
0.7 &360&19\\
BR 1202N & 4.692 & CO & 2--1 & 5.2 & 1.0 & & ? & --- &&7&\\
BR 1202S & 4.695 & CO & 2--1 & 4.6 & 0.8 & & ? & --- &&&&\\
BR 1202 & & CO & 4--3 & 7.6 & 1.5 & 71 & ? & --- & --- && 19$\mu^{-1}$ 
&&\\
TN J0924 & 5.203 & CO & 1--0 & 8.2 & 1.6 & & 1 & 8.2 & & 6.6  &&&\\
SDSS J1148 & 6.419 & CO & 3--2 & 2.6 & 0.6 & 27 & ? & --- & --- & 2.1 & 
2.4$\mu^{-1}$ &&\\
\end{tabular}
\end{table}
\end{landscape}

\begin{table}
\footnotesize
\caption{Dynamical masses and size }
\smallskip
\begin{tabular}{lccccc}\hline\hline
EMG & $M_{\rm gas}\ ^{a}$ & $M_{\rm dyn}$ sin$^{2}i$ $^{b}$ & 
\multicolumn{2}{c}{Source size} & Lens  \\
&& ($=R(\Delta V)^{2}G^{-1}$) & Inferred & Observed & mag.\\
&&& disk diam. & comp. sep. & \\
& ($10^{10}$ \Msun) & ($10^{10}$ \Msun) & (kpc) & (kpc) &  \\ \hline
SMM J02396 & 1.7 & 3.6 & --- & 9.1 & 2.5$^{c}$  \\
F10214 & 0.5 & 0.5 & 0.8 & --- & 17$^{d}$   \\
SMM J16359 & 0.5 & 0.7 & 3.0 & --- & 45$^{e}$ \\
Cloverleaf & 1.9 & 2.5 & 1.5 & --- & 11$^{f}$  \\
SMM J14011 & 0.3--1.5 & 0.3--1.5 & 0.7--7.0 & --- & 2.5--25$^{g}$  \\
SMM J02399 & 4.0 & 32 & 16 & --- & 2.5$^{h}$  \\
4C60.07 & 4+7=11 & 10--43 & --- & 7--30$^{i}$ & 1\\
4C41.17 & 3+3=6 & 6$^{j}$ & --- & 13 & 1\\
APM 08279 & 1.5 & 6.4$^{k}$ & 2.0 & --- & 7$^{j}$  \\
PSS J2322 & 3.9 & 3.0 & 4.0 & --- &  2.5$^{k}$    \\
BRI 1335 & 2.6+3.8=6.4 & 18  & --- & 8.7 &  1 \\
BR 1202S & 4.2+3.7=7.9 & 0.8 & --- & 1.9 &  1 \\
SDSS J1148$^{l}$ & 1.5 & 4.4 & 4.6 & --- &  1\\
SDSS J1148$^{m}$ & 0.5+0.5=1.0 & 1.5 & --- & 1.7 &  1\\
ULIRGs (FWHM)$^{n}$ & &0.7--2.0 & 0.8--2.4 &  --- & 1 \\
ULIRGs (total)$^{p}$ &0.5--1.5 &2--7 & 2.4--6.8 & --- & 1 \\ \hline
\end{tabular}
\end{table}

\begin{table}
\footnotesize
\smallskip
\begin{tabular}{lccccc}
\multicolumn{6}{l}{}\\ \hline
\multicolumn{6}{l}{\scriptsize{$^{a}$From Table 1, where $M_{gas}=\MH2$ 
corrected to include He; $^{b}$R = disk radius; $\Delta V$= FWHM}}\\
\multicolumn{6}{l}{\scriptsize{of line profile or half the velocity 
separation of components; $^{c}$Kneib et al. (1993);}}\\
\multicolumn{6}{l}{\scriptsize{$^d$D Downes \& PM Solomon, manuscript in preparation; $^e$Kneib 
et al. (2004b); $^f$Venturini \& Solomon (2003);}}\\
\multicolumn{6}{l}{\scriptsize{$^g$Downes \& Solomon (2003); $^h$Genzel 
et al. (2004); $^{i}$Range reflects change in component}}\\
\multicolumn{6}{l}{\scriptsize{separation with image tapering; 
$^j$Lewis et al. (2002), authors used HWHM velocity,}}\\
\multicolumn{6}{l}{\scriptsize{yielding $\Mdyn=1.5\times10^{10}$ \Msun\ 
after adjusting for cosmology; $^k$Carilli et al. (2003);}}\\
\multicolumn{6}{l}{\scriptsize{$^l$Low resolution (1.5$\arcsec$) image; 
$^{m}$High resolution (0.15$\arcsec$) image; $^{n}$For radius at half 
maximum }}\\
\multicolumn{6}{l}{\scriptsize{of gas density; $^{p}$For radius of full 
extent of gas.}}\\
\end{tabular}
\end{table}

\cleardoublepage

\begin{table}
\def~{\hphantom{0}}
\footnotesize
\begin{tabular}{llllll} 
\multicolumn{6}{c}{Appendix 1 --- Early (Universe) Molecular (Line Emission) Galaxies} \\ \hline\hline 
\multicolumn{1}{c}{EMG$^{a}$} & \multicolumn{2}{c}{CO coordinates}  & \multicolumn{1}{c}{Redshift} & \multicolumn{1}{c}{Galaxy} & \multicolumn{1}{c}{Lensed?}  \\
 & \multicolumn{1}{c}{R.A. (2000)} & \multicolumn{1}{c}{Decl. (2000)} & \multicolumn{1}{c}{$z$(CO)} & \multicolumn{1}{c}{type} & \multicolumn{1}{c}{($\mu=$ mag.)}  \\ \hline
SMM J02396$^{b}$ & 02:39:56.59 & -01:34:26.6$^{b}$ & 1.062$\pm$0.002$^{b}$ & SMG & $\mu=2.5^{b}$  \\
Q0957+561A$^{c}$ & 10:01:20.88 & +55:53:54.0$^{d}$ & \vlap{1.4141$^{c}$} & \vlap{QSO} & $\mu$=1.6, 1.7$^{d}$  \\
Q0957+561B$^{c}$ & 10:01:21.01 & +55:53:49.4$^{d}$ & & & $\mu$=4.3$^{d}$ \\
HR10$^{e}$ & 16:45:02.26 & +46:26:26.5$^{f}$ & 1.439$\pm$0.001$^{f}$ & ERO & \ \ \ \ ?  \\
IRAS F10214$^{g}$ & 10:24:34.56 & +47:09:09.8$^{h}$ & 2.28581$\pm$0.00005$^{h}$ & QSO & $\mu=17^{h}$  \\
SMM J16371$^{b}$ & 16:37:06.50 & +40:53:13.8$^{b}$ & 2.380$\pm$0.004$^{b}$ & SMG & \ \ \ \ ? \\
SMM J16368$^{i}$ & 16:36:50.43 & +40:57:34.7$^{i}$ & 2.3853$\pm$0.0014$^{i}$ & SMG & \ \ \ \ ? \\
53W002$^{j}$ & 17:14:14.71 & +50:15:30.6$^{k}$ & 2.3927$\pm$0.0003$^{k}$ & Radio & Unlikely \\
SMM J16366$^{b}$ & 16:36:58.23 & +41:05:23.7$^{b}$ & 2.450$\pm$0.002$^{b}$ & SMG & \ \ \ \ ? \\
SMM J04431$^{i}$ & 04:43:07.25 & +02:10:23.3$^{i}$ & 2.5094$\pm$0.0002$^{i}$ & SMG & $\mu=4.4^{l}$ \\
SMM J16359A$^{m}$ & 16:35:54.81 & +66:12:37$^{m}$ & 2.5168$\pm$0.0003$^{m}$ & SMG & $\mu=14^{n}$ \\
SMM J16359B$^{m}$ & 16:35:44.15 & +66:12:24$^{m}$ & 2.5168$\pm$0.0003$^{m}$ & SMG & $\mu=22^{n}$ \\
SMM J16359C$^{m}$ & 16:35:50.85 & +66:12:06$^{m}$ & 2.5168$\pm$0.0003$^{m}$ & SMG & $\mu=9^{n}$ \\
Cloverleaf $^{o}$ & 14:15:45.97 & +11:29:43.2$^{p}$ & 2.5579$\pm$0.0001$^{q}$ & QSO & $\mu=11^{r}$ \\
SMM J14011$^{s}$ & 14:01:04.93 & +02:52:24.1$^{t}$ & 2.5652$\pm$0.0001$^{t}$ & SMG & $\mu$=5--25$^{t}$ \\
VCV J1409$^{u}$ & 14:09:55.50 & +56:28:27.0$^{v}$ & 2.5832$\pm$0.0001$^{v}$ & QSO & \ \ \ \ ? \\
LBQS 0018$^{z}$ & 00:21:27.30 & -02:03:33.0$^{u}$ & 2.620$^{z}$ & QSO & \ \ \ \ ? \\
MG 0414$^{w}$ & 04:15:10.73 & +05:34:41.2$^{x}$ & 2.639$\pm$0.002$^{w}$ & QSO & Yes$^{x}$ \\
MS1512 cB58$^{y}$ & 14:14:22.22 & +36:36:24.8$^{y}$ & 2.7265$\pm$0.0005$^{y}$ & LBG & $\mu=32^{y}$ \\
LBQS 1230$^{aa}$ & 12:33:10.47 & +16:10:53.1$^{bb}$ & 2.741$\pm$0.001$^{aa}$ & QSO & \ \ \ \ ? \\
RX J0911.4$^{u}$ & 09:11:27.50 & +05:50:52.0$^{u}$ & 2.796$\pm$0.001$^{u}$ & QSO & $\mu=22^{cc}$ \\
SMM J02399$^{dd}$ & 02:39:51.89 & $-$01:35:58.8$^{ee}$ & 2.8076$\pm$0.0002$^{ee}$ & SMG & $\mu=2.5^{ee}$ \\
SMM J04135$^{u}$ & 04:13:27.50 & +10:27:40.3$^{u}$ & 2.846$\pm$0.002$^{u}$ & QSO & $\mu=1.3^{ff}$ \\
B3 J2330$^{gg}$ & 23:30:24.84 & +39:27:12.2$^{gg}$ & 3.092$^{gg}$ & Radio & Unlikely \\
SMM J22174$^{b}$ & 22:17:35.20 & +00:15:37.6$^{b}$ &3.099$\pm$0.004$^{b}$ & SMG & \ \ \ \ ? \\
MG 0751$^{hh}$ & 07:51:41.46 & +27:16:31.4$^{hh}$ & 3.200$^{hh}$ & QSO & $\mu=17^{hh}$ \\
SMM J09431$^{i}$ & 09:43:03.74 & +47:00:15.3$^{i}$ & 3.3460$\pm$0.0001$^{i}$ & SMG & $\mu=1.2^{ii}$ \\
SMM J13120$^{b}$ & 13:12:01.20 & +42:42:08.8$^{b}$ & 3.408$\pm$0.002$^{b}$ & SMG & \ \ \ \ ? \\
TN J0121$^{jj}$ & 01:21:42.75 & +13:20:58.0$^{jj}$ & 3.520$^{jj}$ & Radio & \ \ \ \ ? \\
6C1908$^{kk}$  & 19:08:23.70 & +72:20:11.8$^{kk}$ & 3.532$^{kk}$ & Radio & Unlikely \\
4C60.07$^{kk}$ & 05:12:54.75 & +60:30:50.9$^{ll}$ & 3.791$^{kk}$ & Radio & Unlikely \\
4C41.17R$^{mm}$ & 06:50:52.24 & +41:30:31.6$^{mm}$ & 3.7958$\pm$0.0004$^{mm}$ & \vlap{Radio} & \vlap{Unlikely} \\
4C41.17B$^{mm}$ & 06:50:52.12 & +41:30:30.3$^{mm}$ & 3.7888$\pm$0.0008$^{mm}$ & &  \\
APM 08279$^{nn}$ & 08:31:41.70 & +52:45:17.4$^{nn}$ & 3.9114$\pm$0.0002$^{nn}$ & QSO & $\mu=7^{oo}$ \\
PSS J2322$^{pp}$ & 23:22:07.15 & +19:44:22.5$^{qq}$ & 4.1192$\pm$0.0004$^{qq}$ & QSO & $\mu=2.5^{rr}$ \\
BRI 1335N$^{ss}$ & 13.38.03.42 & $-$04:32:34.1$^{tt}$ & 4.4074$\pm$0.0015$^{uu}$ & \vlap{QSO} & \vlap{\ \ \ \ ?} \\
BRI 1335S$^{}$ &13:38:03.40 & $-$04:32:35.4$^{tt}$ & 4.407$^{tt}$ & &  \\
BRI 0952$^{aa}$ & 09:55:00.10 & $-$01:30:07.1$^{aa}$ & 4.4337$\pm$0.0003$^{aa}$ & QSO & $\mu=4^{aa}$ \\
BR 1202N$^{vv}$ & 12:05:22.98 & $-$07:42:29.9$^{tt}$ & 4.6916$^{tt}$ & \vlap{QSO} & \vlap{Likely} \\
BR 1202S$^{vv}$ & 12:05:23.12 & $-$07:42:32.9$^{tt}$ & 4.6947$^{tt}$ & &  \\
TN J0924$^{yy}$ & 09:24:19.92 & -22:01:41.5$^{yy}$ & 5.203$^{yy}$ & Radio & Unlikely \\
SDSS J1148$^{ww}$ & 11:48:16.64 & +52:51:50.3$^{ww}$ & 6.4189$\pm$0.0006$^{xx}$ & QSO & Yes \\ \hline
\end{tabular}
\end{table}

\begin{table} 
\footnotesize
\begin{tabular}{llllll}\hline
\multicolumn{6}{l}{\scriptsize{$^{a}$Reference is to discovery paper; $^{b}$Greve et al. (2004b); $^{c}$Planesas et al. (1999); }}\\
\multicolumn{6}{l}{\scriptsize{$^{d}$Krips et al. (2004); $^{e}$Andreani et al. (2000); $^{f}$Greve, Ivison \& Papadopoulos (2003);}}\\
\multicolumn{6}{l}{\scriptsize{$^{g}$Brown \& Vanden Bout (1991), Solomon, Downes \& Radford (1992a); $^{h}$Downes \& Solomon (2004); }}\\
\multicolumn{6}{l}{\scriptsize{$^{i}$Neri et al. (2003); $^{j}$Scoville et al. (1997); $^{k}$Alloin, Barvainis \& Guilloteau (2000); }}\\
\multicolumn{6}{l}{\scriptsize{$^{l}$Smail et al. (1999); $^{m}$Sheth et al. (2004);$^{n}$Kneib et al. (2004b); $^{o}$Barvainis et al. (1994); }}\\
\multicolumn{6}{l}{\scriptsize{$^{p}$Center of four lensed components, Kneib et al. (1998); $^{q}$Barvainis et al. (1997); }}\\
\multicolumn{6}{l}{\scriptsize{$^{r}$Venturini \& Solomon (2003); $^{s}$Frayer et al. (1999); $^{t}$Downes \& Solomon (2003); }}\\
\multicolumn{6}{l}{\scriptsize{$^{u}$Hainline et al. (2004); $^{v}$Beelen et al. (2004); $^{w}$Barvainis et al. (1998);}}\\
\multicolumn{6}{l}{\scriptsize{$^{x}$Hewitt et al. (1992); $^{y}$Baker et al. (2004); $^{z}$K Izaak, private communication; }}\\
\multicolumn{6}{l}{\scriptsize{$^{aa}$Guilloteau et al. (1999); $^{bb}$Hewett et al. (1995); $^{cc}$Barvainis \& Ivison (2002a);   }}\\
\multicolumn{6}{l}{\scriptsize{$^{dd}$Frayer et al. (1998); $^{ee}$Genzel et al. (2003); $^{ff}$Knudsen et al. (2003);  $^{gg}$De Breuck et al. (2003b); }}\\
\multicolumn{6}{l}{\scriptsize{$^{hh}$Barvainis, Alloin \& Bremer (2002); $^{ii}$Cowie, Barger \& Kneib (2002); }}\\
\multicolumn{6}{l}{\scriptsize{$^{jj}$De Breuck, Neri \& Omont (2003a); $^{kk}$Papadopoulos et al. (2000); $^{ll}$Greve, Ivison \& Papadopoulos (2004); }}\\
\multicolumn{6}{l}{\scriptsize{$^{mm}$De Breuck et al. (2004); $^{nn}$Downes et al. (1999); $^{oo}$Lewis et al. (2002); $^{pp}$Cox et al. (2002);  }}\\
\multicolumn{6}{l}{\scriptsize{$^{qq}$Carilli et al. (2002b); $^{rr}$Carilli et al. (2003); $^{ss}$Guilloteau et al. (1997); $^{tt}$Carilli et al. (2002b);   }}\\
\multicolumn{6}{l}{\scriptsize{$^{uu}$Carilli, Menten \& Yun (1999); $^{vv}$Omont et al. (1996b) \& Ohta et al. (1995); $^{ww}$Walter et al. (2003); }}\\
\multicolumn{6}{l}{\scriptsize{$^{xx}$Bertoldi et al. (2003b); $^{yy}$Klamer et al. (2005). }}\\
 \hline
\end{tabular}
\end{table}

\begin{table}
\footnotesize
\def~{\hphantom{0}}
\smallskip
\begin{tabular}{lr@{ }lr@{$\pm$}lr@{$\pm$}lr@{$\pm$}lr@{$\pm$}lccr@{.}l}\hline\hline
\multicolumn{1}{c}{EMG} & \multicolumn{2}{c}{Transition} & \multicolumn{2}{c}{$S\Delta v$} & \multicolumn{2}{c}{$\Delta v$} &       \multicolumn{2}{c}{$S(\rm peak)$} & \multicolumn{2}{c}{$L^{\prime}(\rm app.)$} & \multicolumn{1}{c}{$L^{\prime}(\rm int.)$} &  \multicolumn{3}{c}{$M_{\rm gas}$} \\
& & & \multicolumn{2}{c}{(Jy km s$^{-1}$)} & \multicolumn{2}{c}{(km s$^{-1}$)} & \multicolumn{2}{c}{(mJy)} & \multicolumn{2}{c}{(10$^{10}$L$'_*)^a$} & \multicolumn{1}{c}{(10$^{10}$L$'_*)^a$} &  \multicolumn{3}{c}{(10$^{10}$\Msun)}\\ \hline
SMM J02396 & CO & 2--1$^{b}$ & 3.4 & 0.3 & 780 & 60 & \multicolumn{2}{c}{$\sim{5}$} & 5.1 & 0.5 & 2.0 &  &  1 & 6 \\
Q0957+561A(r) & CO & 2--1$^{c}$ & 0.34 & 0.06 & 160 & 20 & 2.1 & 0.2 & 0.9 & 0.2 & \multicolumn{1}{c}{0.6} & \multirow{3}{*}{$\left. \phantom{\matrix{ \cr x \cr x \cr x }} \right\}$}  & \multicolumn{2}{c}{ }\\
Q0957+561A(b) & CO & 2--1$^{c}$ & 0.25 & 0.06 & 280 & 60 & 0.9 & 0.2 & 0.7 & 0.2 & 0.4 & & 0 & 4 \\
Q0957+561B    & CO & 2--1$^{c}$ & 0.61 & 0.06 & 280 & 50 & 2.2 & 0.2 & 1.6 & 0.2 & 0.4  & & \multicolumn{2}{c}{ } \\
HR10  & CO & 1--0$^{d}$ & 0.6 & 0.1 & \multicolumn{2}{c}{---} & \multicolumn{2}{c}{$\sim{0.7}$} & 6.5 & 1.1 & \multicolumn{1}{c}{---} & & 5  & 2$\mu^{-1}$ \\
 & CO & 2--1$^{e}$ & \multicolumn{2}{c}{1.45} & \multicolumn{2}{c}{400} & \multicolumn{2}{c}{$\sim{4}$} &  \multicolumn{2}{c}{3.8} & \multicolumn{1}{c}{---} & & \multicolumn{2}{c}{ }\\ 
& CO & 5--4$^{e}$ & \multicolumn{2}{c}{1.35} & \multicolumn{2}{c}{380} & \multicolumn{2}{c}{$\sim{7}$} &  \multicolumn{2}{c}{0.6} & \multicolumn{1}{c}{---}& & \multicolumn{2}{c}{ }\\
IRAS F10214 & CO & 3--2$^{f}$ & 4.1 & 0.6 & 220 & 20 & 14.5 & 1.5 & 11.3 & 1.7  & 0.7  & & 0 & 6  \\
& & 4--3$^{f}$ & 5.5 & 1.0 & 220 & 40 & 23 & 4 & 8.6 & 0.16 & 0.5  & & \multicolumn{2}{c}{ } \\
& & 6--5$^{f}$ & 8.5 & 2.0 & 200 & 30 & 32 & 6 & 5.9 & 1.4 & 0.4 & & \multicolumn{2}{c}{ }\\
& & 7--6$^{f}$ & 7.1 & 2.0   & 210 & 40 & 19 & 5 & 3.6 & 1.0 & \multicolumn{1}{c}{0.2} & & \multicolumn{2}{c}{ }\\
& HCN & 1--0$^{g}$ & 0.05 & 0.01 & 140 & 30 & 0.45 & 0.08 & 2.3 & 0.4 & 0.14  & & 1 & 0 \\
& \TCI & 1--0$^{h}$ &  1.6 & 0.2 & 160 & 30 & 9.2 & 1.0 & 2.1 & 0.3 & \multicolumn{1}{c}{---}& & \multicolumn{2}{c}{ } \\
SMM J16371 & CO & 3--2$^{b}$ & 1.0 & 0.2 & 830 & 130 & \multicolumn{2}{c}{$\sim 1$} & 3.0 & 0.6 & \multicolumn{1}{c}{---} & & 2 & 4$\mu^{-1}$\\
SMM J16368 & CO & 3--2$^{i}$ & 2.3 & 0.2 & 840 & 110 & \multicolumn{2}{c}{$\sim 3$} &6.9 & 0.6 & \multicolumn{1}{c}{---}& & 5 & 5$\mu^{-1}$ \\
53W002 & CO & 3--2$^{j}$ & 1.20 & 0.15 & 420 & 40 & 2.5 & 0.8 & 3.6 & 0.4 & \multicolumn{1}{c}{3.6}& & 2 & 9\\
SMM J16366 & CO & 3--2$^{b}$ & 1.8 & 0.3 & 870 & 80 & \multicolumn{2}{c}{$\sim 2$} & 5.6 & 0.9 & \multicolumn{1}{c}{---}& & 4 & 5$\mu^{-1}$ \\
SMM J04431 & CO & 3--2$^{i}$ & 1.4 & 0.2 & 350 & 60 & 3.5 & 0.5 & 4.5 & 0.6 & 1.0  & & 0 & 8\\
SMM J16359A & CO & 3--2$^{k}$ & 1.67 & 0.13 & \multicolumn{2}{c}{$\sim 500$} &  \multicolumn{2}{c}{$\sim 4$}  & 5.5 & 0.4   &  0.4  & \multirow{3}{*}{$\left. \phantom{\matrix{ \cr x \cr x \cr x }} \right\}$} & \multicolumn{2}{c}{ } \\
SMM J16359B & CO & 3--2$^{k}$ & 2.50 & 0.12 & \multicolumn{2}{c}{$\sim 500$} &  \multicolumn{2}{c}{$\sim 7$}  & 8.2 & 0.4 &   0.4  & & 0 & 4\\
SMM J16359C & CO & 3--2$^{k}$ & 1.58 & 0.17 & \multicolumn{2}{c}{$\sim 500$} &  \multicolumn{2}{c}{$\sim 4$}  & 5.2 & 0.6  & 0.6  & & \multicolumn{2}{c}{ } \\
\end{tabular}
\end{table}

\begin{table}
\footnotesize
\def~{\hphantom{0}}
\smallskip
\begin{tabular}{lr@{ }lr@{$\pm$}lr@{$\pm$}lr@{$\pm$}lr@{$\pm$}lccr@{.}l}\hline\hline
\multicolumn{1}{c}{EMG} & \multicolumn{2}{c}{Transition} & \multicolumn{2}{c}{$S\Delta v$} & \multicolumn{2}{c}{$\Delta v$} &       \multicolumn{2}{c}{$S(\rm peak)$} & \multicolumn{2}{c}{$L^{\prime}(\rm app.)$} & \multicolumn{1}{c}{$L^{\prime}(\rm int.)$} &  \multicolumn{3}{c}{$M_{\rm gas}$} \\
& & & \multicolumn{2}{c}{(Jy km s$^{-1}$)} & \multicolumn{2}{c}{(km s$^{-1}$)} & \multicolumn{2}{c}{(mJy)} & \multicolumn{2}{c}{(10$^{10}$L$'_*)^a$} & \multicolumn{1}{c}{(10$^{10}$L$'_*)^a$} &  \multicolumn{3}{c}{(10$^{10}$\Msun)}\\ \hline
Cloverleaf & CO & 3--2$^{l}$ & 13.2 & 0.2 & 416 & 6 & 30.0 & 1.7 & 44 & 1 & 4.0  & & 3 & 2 \\
& & 4--3$^{m}$ & 21.1 & 0.8 & 375 & 16 & 53 & 2 & 40 & 2 & 3.6  & & \multicolumn{2}{c}{ }\\
& & 5--4$^{m}$ & 24.0 & 1.4 & 398 & 25 & 56 & 3 & 29 & 2 & 2.6  & & \multicolumn{2}{c}{ }\\
& & 7--6$^{n}$ & 36 & 6 & \multicolumn{2}{c}{$\sim 450$} & 80 & 8 & 22 & 4 & 2.0  & & \multicolumn{2}{c}{ }\\
& HCN & 1--0$^{o}$ & 0.069 & 0.012 & \multicolumn{2}{c}{$\sim 300$} & 0.24 & 0.04 & 3.5 & 0.6 & 0.32  & & 2 & 2\\
& \TCI & 1--0$^{h}$ & 3.9 & 0.6 & 360 & 60 & 11.2 & 2.0 & 6.5 & 1.0 & \multicolumn{1}{c}{---} & & \multicolumn{2}{c}{ }\\
& & 2--1$^{l}$ & 5.2 & 0.3 & 468 & 25 & 13.2 & 2.9 & 3.2 & 0.2 & ---  & &\multicolumn{2}{c}{ }\\
SMM J14011 & CO & 3--2$^{p}$ & 2.8 & 0.3 & 190 & 11 & 13.2 & 1.0 & 9.4 & 1.0 & \multicolumn{1}{c}{0.4--1.9}  & \multicolumn{3}{c}{0.3--1.5}\\
& & 7--6$^{p}$ & 3.2 & 0.5 & 170 & 30 & 12.4 & 3.0 & 2.0 & 0.3 & \multicolumn{1}{c}{0.08--0.4}  & \multicolumn{2}{c}{ } \\
& \TCI & 1--0$^{h}$ & 1.8 & 0.3 & 235 & 45 & 7.3 & 1.5 & 3.0 & 0.5 & \multicolumn{1}{c}{---} & & \multicolumn{2}{c}{ }\\
VCV J1409 & CO & 3--2$^{q}$ & 2.3 & 0.2 & 311 & 28 & 6 & 1 & 7.9 & 0.7 & \multicolumn{1}{c}{---} & & 6 & 3$\mu^{-1}$\\
& & 7--6$^{q}$ & 4.1 & 1.0 & \multicolumn{2}{c}{$\sim 300$} & 10 & 3 & 2.6 & 0.6 & \multicolumn{1}{c}{---} & & \multicolumn{2}{c}{ } \\
& HCN & 1--0$^{r}$ & 0.007 & 0.002 & \multicolumn{2}{c}{$\sim200$} & 0.08 & 0.03 & 0.7 & 0.2 & \multicolumn{1}{c}{---}  & & 4 & 9$\mu^{-1}$\\
LBQS 0018 & CO & 3--2$^{s}$ & 1.55 & 0.26 & 163 & 29 & \multicolumn{2}{c}{---} & 5.4 & 0.9 & --- & & 4 & 3$\mu^{-1}$ \\
MG0414 & CO & 3--2$^{t}$ & 2.6 & 0.5 & \multicolumn{2}{c}{$\sim 580$} & \multicolumn{2}{c}{4.4} & \multicolumn{2}{c}{9.2}  & \multicolumn{1}{c}{---} & & 7 & 4$\mu^{-1}$\\
MS1512-cB58 & CO & 3--2$^{u}$ & 0.37 & 0.08 & 175 & 45 & \multicolumn{2}{c}{$\sim2$} & 1.4 & 0.3 & 0.043  & & 0 & 03\\
LBQS 1230 & CO & 3--2$^{v}$ & 0.80 & 0.26 & \multicolumn{2}{c}{---} & \multicolumn{2}{c}{---} & 3.0 & 1.0 & \multicolumn{1}{c}{---} & & 2 & 4$\mu^{-1}$\\
RX J0911.4 & CO & 3--2$^{w}$ & 2.9 & 1.1 & 350 & 60 & \multicolumn{2}{c}{$\sim 8$} & 11.3 & 4.3 & 0.52  & & 0 & 4\\
SMM J02399 & CO & 3--2$^{x}$ & 3.1 & 0.4 & \multicolumn{2}{c}{$\sim 1100$} & \multicolumn{2}{c}{$\sim 4$} & 12.2 & 1.6 & 4.9  & & 3 & 9\\
SMM J04135 & CO & 3--2$^{w}$ & 5.4 & 1.3 & 340 & 120 & \multicolumn{2}{c}{$\sim 16$} & 22 & 5 & 17  & & 13 & 0\\
B3 J2330 & CO & 4--3$^{y}$ & 1.3 & 0.3 & \multicolumn{2}{c}{$\sim 500$} & \multicolumn{2}{c}{2.5} & 3.4 & 0.8 & \multicolumn{1}{c}{3.4} & & 2 & 7\\
SMM J22174 & CO & 3--2$^{b}$ & 0.8 & 0.2 & 780 & 100 & \multicolumn{2}{c}{$\sim 1$} & 3.7 & 0.9 & \multicolumn{1}{c}{---} & & 3 & 0\\
MG0751 & CO & 4--3$^{z}$ & 5.96 & 0.45 & 390 & 40 & \multicolumn{2}{c}{$\sim 15$} & 16 & 1 & 1.0  & & 0 & 8\\
SMM J09431 & CO & 4--3$^{i}$ & 1.1 & 0.1 & 420 & 50 & 2.5 & 0.5 & 3.2 & 0.3 & 2.7  & & 2 & 2\\
SMM J13120 & CO & 4--3$^{b}$ & 1.7 & 0.3 & 530 & 50 & \multicolumn{2}{c}{$\sim 3$} & 5.2 & 0.9 & \multicolumn{1}{c}{---} & & 4 & 2$\mu^{-1}$\\
TN J0121 & CO & 4--3$^{aa}$ & 1.2 & 0.4 & \multicolumn{2}{c}{$\sim 700$} & \multicolumn{2}{c}{$\sim 2$} & 5.4 & 1.0 & \multicolumn{1}{c}{5.4} & & 4 & 3\\
6C1908 & CO & 4--3$^{bb}$ & 1.62 & 0.30 & 530 & 70 & \multicolumn{2}{c}{$\sim 3$} & 5.2 & 1.0 & \multicolumn{1}{c}{5.2} & & 4 & 2\\
4C60.07 & CO & 1--0$^{cc}$ & 0.15 & 0.03 & \multicolumn{2}{c}{$\sim 550$} & 0.27 & 0.05 & 8.7 & 1.7 & 8.7 & & 7 & 0\\
& CO & 1--0$^{cc}$ & 0.09 & 0.01 & 165 & 24 & 0.30 & 0.10 & 5.2 & 0.6 & 5.2 & & 4 & 2\\
& & 4--3$^{bb}$ & 1.65 & 0.35 & \multicolumn{2}{c}{$\sim550$} & \multicolumn{2}{c}{$\sim3$} & 6.0 & 0.9 & 6.0 & & \multicolumn{2}{c}{ }\\
& & 4--3$^{bb}$ & 0.85 & 0.2 & \multicolumn{2}{c}{$\sim150$} & \multicolumn{2}{c}{$\sim6$} & 3.0 & 0.2 & 3.0 & & \multicolumn{2}{c}{ }\\
4C41.17R & CO & 4--3$^{dd}$ & 1.20 & 0.15 & 500 & 100 & \multicolumn{2}{c}{$\sim2.5$} & 4.3 & 0.5  & \multicolumn{1}{c}{4.3} & \multirow{2}{*}{$\left. \phantom{\matrix{ \cr x   }} \right\}$}  & \omit  \multirow{2}{*}{5.2}\\
4C41.17B & CO & 4--3$^{dd}$ & 0.60 & 0.15 & 500 & 150 & \multicolumn{2}{c}{$\sim1.5$} &  2.2 & 0.5 & \multicolumn{1}{c}{2.2} &  \multicolumn{3}{c}{ } \\
APM 08279 & CO & 1--0$^{ee}$ & 0.150 & 0.045 & \multicolumn{2}{c}{---} & \multicolumn{2}{c}{---} & 9.1 & 2.7 & 1.3  & & 1 & 0 \\
& & 4--3$^{ff}$ & 3.7 & 0.5 & 480 & 35 & 7.4 & 1.0 & 14 & 2 & 2.0  & & \multicolumn{2}{c}{ }\\
& & 9--8$^{ff}$ & 9.1 & 0.8 & \multicolumn{2}{c}{$\sim 500$} & 17.9 & 1.4 & 6.8 & 0.6 & 1.0  & \multicolumn{2}{c}{ }\\
(N/NE comp.)$^{gg}$ & & 2--1$^{ee}$ & 1.15 & 0.54 & \multicolumn{2}{c}{---} & \multicolumn{2}{c}{---} & 17 & 8 & \multicolumn{1}{c}{---} & \multicolumn{2}{c}{ }\\
PSS J2322 & CO & 1--0$^{hh}$ & 0.19 & 0.08 & 200 & 70 & 0.9 & 0.2 & 12 & 5 & 5.0  & & 4 & 0\\
& & 2--1$^{hh}$ & 0.92 & 0.30 & \multicolumn{2}{c}{---} & 2.70 & 0.24 & 15 & 5 & 6.1  & & \multicolumn{2}{c}{ }\\
& & 4--3$^{ii}$ & 4.21 & 0.40 & 375 & 40 & \multicolumn{2}{c}{10.5} & 17.3 & 1.6 & 6.9  & & \multicolumn{2}{c}{ }\\
& & 5--4$^{ii}$ & 3.74 & 0.56 & 275 & 50 & \multicolumn{2}{c}{12} & 9.8 & 1.5 & 3.9  & & \multicolumn{2}{c}{ }\\
& \TCI & 1--0$^{jj}$ & 0.81 & 0.12 & 319 & 66 & \multicolumn{2}{c}{2.4} & 3.3 & 0.5 & \multicolumn{1}{c}{---} & & \multicolumn{2}{c}{ }\\
BRI 1335N & CO & 2--1$^{kk}$ & 0.18 & 0.06 & \multicolumn{2}{c}{---} & 0.45 & 0.14 & 3.3 & 1.1 & \multicolumn{1}{c}{---} & & 2 & 6$\mu^{-1}$\\
BRI 1335S & CO & 2--1$^{kk}$ & 0.26 & 0.06 & \multicolumn{2}{c}{---} & 0.67 & 0.14 & 4.8 & 1.1 & \multicolumn{1}{c}{---} & & 3 & 8$\mu^{-1}$\\
BRI 1335 & CO & 5--4$^{ll}$ & 2.8 & 0.3 & 420 & 60 & 6 & 1 & 8.2 & 0.9 & \multicolumn{1}{c}{---} & & 6 & 6$\mu^{-1}$\\
BRI 0952 & CO & 5--4$^{v}$ & 0.91 & 0.11 & \multicolumn{2}{c}{230}  & \multicolumn{2}{c}{$\sim3$} & 2.8 & 0.3 & 0.7  & & 0 & 5\\
\end{tabular}
\end{table}

\begin{table}
\footnotesize
\def~{\hphantom{0}}
\smallskip
\begin{tabular}{lr@{ }lr@{$\pm$}lr@{$\pm$}lr@{$\pm$}lr@{$\pm$}lccr@{.}l}\hline\hline
\multicolumn{1}{c}{EMG} & \multicolumn{2}{c}{Transition} & \multicolumn{2}{c}{$S\Delta v$} & \multicolumn{2}{c}{$\Delta v$} &       \multicolumn{2}{c}{$S(\rm peak)$} & \multicolumn{2}{c}{$L^{\prime}(\rm app.)$} & \multicolumn{1}{c}{$L^{\prime}(\rm int.)$} &  \multicolumn{3}{c}{$M_{\rm gas}$} \\
& & & \multicolumn{2}{c}{(Jy km s$^{-1}$)} & \multicolumn{2}{c}{(km s$^{-1}$)} & \multicolumn{2}{c}{(mJy)} & \multicolumn{2}{c}{(10$^{10}$L$'_*)^a$} & \multicolumn{1}{c}{(10$^{10}$L$'_*)^a$} &  \multicolumn{3}{c}{(10$^{10}$\Msun)}\\ \hline
BR 1202N & CO & 2--1$^{kk}$ & 0.26 & 0.05 & \multicolumn{2}{c}{---} & 0.44 & 0.07 & 5.2 & 1.0 & \multicolumn{1}{c}{---} & & 4 & 2\\
& & 5--4$^{mm}$ & 1.3 & 0.3 & \multicolumn{2}{c}{$\sim 350$} & \multicolumn{2}{c}{$\sim3$} & 4.2 & 1.0 & \multicolumn{1}{c}{---} & &  \multicolumn{2}{c}{ } \\
BR 1202S & CO & 2--1$^{kk}$ & 0.23 & 0.04 & \multicolumn{2}{c}{---} & 0.77 & 0.10 & 4.6 & 0.8 & \multicolumn{1}{c}{---} &  & 3 & 7\\
& & 5--4$^{mm}$ & 1.1 & 0.2 & \multicolumn{2}{c}{$\sim 190$} & \multicolumn{2}{c}{$\sim5$} & 3.5 & 0.6 & \multicolumn{1}{c}{---} & & \multicolumn{2}{c}{ } \\
BR 1202 & CO & 4--3$^{mm}$ & 1.5 & 0.3 & \multicolumn{2}{c}{---} & \multicolumn{2}{c}{---} & 7.6 & 1.5 & \multicolumn{1}{c}{---} & & 6 & 1\\
& & 7--6$^{mm}$ & 3.1 & 0.9 & \multicolumn{2}{c}{$\sim 275$} & \multicolumn{2}{c}{$\sim10$}  & 5.1 & 1.5 & \multicolumn{1}{c}{---} & & \multicolumn{2}{c}{ }\\
TN J0924 & CO & 1--0$^{}$ & 0.087 & 0.017 & \multicolumn{2}{c}{$\sim 300$} & 0.52 & 0.12 & 8.2 & 1.6 & 8.2 & & 6 & 6   \\
& & 5--4$^{}$ & 1.19 & 0.27 & \multicolumn{2}{c}{$\sim 300$} & 7.8 & 2.7 & 4.5 & 1.0 & 4.5 & &  \multicolumn{2}{c}{ }  \\
SDSS J1148 & CO & 3--2$^{nn}$ & 0.18 & 0.04 & \multicolumn{2}{c}{$\sim 250$} & \multicolumn{2}{c}{$\sim0.6$} & 2.6 & 0.6 & \multicolumn{1}{c}{---} & & 2 & 1${\mu^{-1}}$\\
& & 6--5$^{oo}$ & 0.73 & 0.076 & \multicolumn{2}{c}{$\sim 280$} & \multicolumn{2}{c}{$\sim 2.5$} & 2.6 & 0.3 & \multicolumn{1}{c}{---} & & \multicolumn{2}{c}{ } \\
& & 7--6$^{oo}$ & 0.640 & 0.088 & \multicolumn{2}{c}{$\sim 280$} & \multicolumn{2}{c}{$\sim 2.1$} & 1.7 & 0.2 & \multicolumn{1}{c}{---} & & \multicolumn{2}{c}{ }\\ \hline
\end{tabular}
\end{table}

\begin{table}
\footnotesize
\def~{\hphantom{0}}
\smallskip
\begin{tabular}{lr@{ }lr@{$\pm$}lr@{$\pm$}lr@{$\pm$}lr@{$\pm$}lr@{$\pm$}l} \hline
\multicolumn{12}{l}{\scriptsize{$^{a}L'_*$=K km s$^{-1}$ pc$^{2}$; $^{b}$Greve et al. (2004b); $^{c}$Krips et al. (2003); $^{d}$Greve, Ivison \& Papadopoulos (2003); }}\\
\multicolumn{12}{l}{\scriptsize{$^{e}$Andreani et al. (2000); $^{f}$D Downes \& PM Solomon, manuscript in preparation; }}\\
\multicolumn{12}{l}{\scriptsize{$^{g}$Vanden Bout, Solomon \& Maddalena (2004); $^{h}$Wei\ss\ et al. (2005);  $^{i}$Neri et al. (2003);}}\\
\multicolumn{12}{l}{\scriptsize{$^{j}$Alloin, Barvainis \& Guilloteau (2000); $^{k}$Kneib et al. (2004b); $^{l}$Wei\ss\  et al. (2003); }}\\
\multicolumn{12}{l}{\scriptsize{$^{m}$Barvainis et al. (1997); $^{n}$Kneib et al. (1998); $^{o}$Solomon et al. (2003); $^{p}$Downes \& Solomon (2003);}}\\
\multicolumn{12}{l}{\scriptsize{ $^{q}$Beelen et al. (2004); $^{r}$Carilli et al. (2004); $^{s}$K Izaak, private communication; }}\\
\multicolumn{12}{l}{\scriptsize{$^{t}$Barvainis et al. (1998); $^{u}$Baker et al. (2004); $^{v}$Guilloteau et al. (1999); $^{w}$Hainline et al. (2004); }}\\
\multicolumn{12}{l}{\scriptsize{$^{x}$Genzel et al. (2003); $^{y}$De Breuck et al. (2003a); $^{z}$Barvainis, Alloin \& Bremer (2002); }}\\
\multicolumn{12}{l}{\scriptsize{$^{aa}$De Breuck et al. (2003b); $^{bb}$Papadopoulos et al. (2000); $^{cc}$Greve, Ivison \& Papadopoulos (2004); }}\\
\multicolumn{12}{l}{\scriptsize{$^{dd}$De Breuck et al. (2004); $^{ee}$Papadopoulos et al. (2001); $^{ff}$Downes et al. (1999);}}\\

\multicolumn{12}{l}{\scriptsize{$^{gg}$Components lie 2--3$^{\prime\prime}$to  N and NE and may be unrelated to the nuclear source;}}\\
\multicolumn{12}{l}{\scriptsize{$^{hh}$Carilli et al. (2002b); $^{ii}$Cox et al. (2002); $^{jj}$Pety et al. (2004); $^{kk}$Carilli et al. (2002b); }}\\
\multicolumn{12}{l}{\scriptsize{$^{ll}$Guilloteau et al. (1997); $^{mm}$Omont et al. (1996b); $^{nn}$Walter et al. (2003); $^{oo}$Bertoldi et al. (2003b).}} \\
\multicolumn{12}{l}{\scriptsize{}}\\ 
\end{tabular}
\end{table}

\begin{table}
\footnotesize
\smallskip
\begin{tabular}{lr@{\tiny }llllcc}\hline\hline
\multicolumn{1}{c}{EMG} & \multicolumn{2}{c}{Band} & \multicolumn{1}{c}{Flux density} & \multicolumn{2}{c}{$L_{\rm FIR}(\rm app.)$} & \multicolumn{1}{c}{$L_{\rm FIR}(\rm int.)$}& $M_{\rm dust}$ \\
& \multicolumn{2}{c}{$(\mu \rm m)$} & \multicolumn{1}{c}{(mJy)} & \multicolumn{2}{c}{(10$^{12}$ L$_{\odot}$)} & \multicolumn{1}{c}{(10$^{12}$ L$_{\odot}$)} & ($10^{8}$ M$_{\odot}$) \\ \hline
SMM02396 & 850 &$^{a}$ & 11 & \multirow{2}{*}{$\left. \phantom{\matrix{ \cr x \cr x  }} \right\}$}& \multirow{2}{*}{16.3$^{a}$} & \multirow{2}{*}{6.5$^{a}$} & \\
& 450 &$^{a}$ & 42 & & &  \\
Q0957+561 & 850 &$^{b}$ & 7.5$\pm$1.4 & & 14$^{c}$  & 6$^{c}$ & 2.5$^{b}$ \\
HR10 & 1350 &$^{d}$ & 2.13$\pm$0.63 & \multirow{3}{*}{$\left. \phantom{\matrix{ \cr x \cr x \cr x }} \right\}$} & &  & \\
& 850 &$^{d}$ & 4.89$\pm$0.74 & & 6.5$^{d}$ & & 6.8$\mu^{-1}\ ^{d}$  \\
& 450 &$^{d}$ & 32.3$\pm$8.5 & & & &  \\
& 850 &$^{e}$ & 8$\pm$2 & & 9$^{e}$ & & 9$\mu^{-1}\ ^{e}$ \\
IRAS F10214 & 1410 &$^{f}$ & 5.7$\pm$1.0 & \multirow{7}{*}{$\left. \phantom{\matrix{ \cr x \cr x \cr x \cr x \cr x \cr x \cr x }} \right\}$} &  & & 0.23$^{f}$\\
& 1240 &$^{f}$ & 10$\pm$2 & & &  & \\
& 1230 &$^{g}$ & 9.6$\pm$1.4 & & & &  \\
& 1100 &$^{h}$ & 24$\pm$5 & & 60$^{f}$ &  3.6$^{f}$ & \\
& 850 &$^{h}$ & 50$\pm$5 & & & & \\
& 450 &$^{i}$ & 273$\pm$45 & & &  & \\
& 350 &$^{j}$ & 383$\pm$51 & & &  & \\
SMM J16371 & 1300 &$^{k}$ & 4.2$\pm$1.1 & & & &  \\
& 850 &$^{l}$ & 11.2$\pm$2.9 & & & & \\
SMM J16368 & 1300 &$^{m}$ & 2.5$\pm$0.4 &  &  & & \\
& 850 &$^{n}$ & 8.2$\pm$1.7 &  & 16$^{c}$ & & \\
53W002 & 1300 &$^{o}$ & 1.7$\pm$0.4 & & &  & \\
SMM J16366 & 850 &$^{n}$ & 10.7$\pm$2.0 &  & 20$^{c}$ & & \\
SMM J04431 & 1300 &$^{l}$ & 1.1$\pm$0.3 & & &  & \\
& 850 &$^{p}$ & 7.2$\pm$1.7 & & 13$^{m}$ &  3$^{m}$ & \\
SMM J16359 & 1350 &$^{q}$ & 3.0$\pm$0.7 & & & &  \\
SMM J16359A & 850 &$^{r}$ & 11$\pm$1 & \multirow{6}{*}{$\left. \phantom{\matrix{ \cr x \cr x \cr x \cr x \cr x \cr x}} \right\}$}& \multirow{6}{*}{45$^{r}$} & \multirow{6}{*}{1$^{r}$} & \multirow{6}{*}{2$^{s}$}\\
& 450 &$^{r}$ & 45$\pm$9 & & &  & \\
SMM J16359B & 850 &$^{r}$ & 17$\pm$2 & & & &  \\
& 450 &$^{r}$ & 75$\pm$15 & & & & \\
SMM J16359C & 850 &$^{r}$ & 9$\pm$1 & & &  & \\
& 450 &$^{r}$ & 32$\pm$6 & & & &  \\
Cloverleaf & 1300 &$^{b}$ & 18$\pm$2 & \multirow{3}{*}{$\left. \phantom{\matrix{ \cr x \cr x \cr x }} \right\}$} &  &  & \\
& 850 &$^{b}$ & 58.8$\pm$8.1 & & 59$^{t}$ &  5.4$^{t}$ & 1.5$^{t}$ \\
& 450 &$^{b}$ & 224$\pm$38 & & & & \\
& 350 &$^{j}$ & 293$\pm$14 & & 77$^{j}$ &  7$^{j}$ & 3.5$^{j}$\\
SMM J14011 & 1350 &$^{u}$ & 2.5$\pm$0.8 &  & 20$^{u}$ &  0.8--4.0$^{u}$ & 0.13--0.65$^{v}$ \\
& 850 &$^{w}$ & 14.6$\pm$1.8 & & & &  \\
& 450 &$^{w}$ & 41.9$\pm$6.9 & & & & \\
VCV J1409 & 1300 &$^{x}$ & 10.7$\pm$0.6 &  & 43$^{x}$  & & \\
& 350 &$^{y}$ &  $159\pm14^{y}$& & 35$^{y}$  &   & 38$\mu^{-1}\ ^{x}$\\
LBQS 0018 & 850 &$^{z}$ & 17.2$\pm$2.9 & & 33$^{c}$ & \\
MG0414 & 3000 &$^{b}$ & 40$\pm$2 & & & &  \\
& 1300 &$^{b}$ & 20.7$\pm$1.3 & & & &   \\
& 850 &$^{b}$ & 16.7$\pm$3.8 &  & 32$^{c}$  &  & \\
& 450 &$^{b}$ & 66$\pm$16 & & & &  \\
MS1512--cB58 & 1200 &$^{aa}$ & 1.06$\pm$0.35 & &  & & \\
& 850 &$^{bb}$ & 4.2$\pm$0.9 &  & 3.1$^{bb}$ &  0.1$^{bb}$ & \\
\end{tabular}
\end{table}

\begin{table} 
\footnotesize
\smallskip
\begin{tabular}{lr@{\tiny }llllcc}\hline\hline
\multicolumn{1}{c}{EMG} & \multicolumn{2}{c}{Band} & \multicolumn{1}{c}{Flux density} & \multicolumn{2}{c}{$L_{\rm FIR}(\rm app.)$} & \multicolumn{1}{c}{$L_{\rm FIR}(\rm int.)$}& $M_{\rm dust}$ \\
& \multicolumn{2}{c}{$(\mu \rm m)$} & \multicolumn{1}{c}{(mJy)} & \multicolumn{2}{c}{(10$^{12}$ L$_{\odot}$)} & \multicolumn{1}{c}{(10$^{12}$ L$_{\odot}$)} & ($10^{8}$ M$_{\odot}$) \\ \hline
LBQS 1230 & 1350 &$^{cc}$ & 3.3$\pm$0.5 & & & &  \\
& 1250 &$^{cc}$ & 7.5$\pm$1.4 & & & &  \\
& 350 &$^{j}$ & 104$\pm$21 & & 36$^{j}$ &  & 11$\mu^{-1}\ ^{j}$\\
RX J0911.4 & 3000 &$^{b}$ & 1.7$\pm$0.3 & & & & \\
& 1300 &$^{b}$ & 10.2$\pm$1.8 & & & & \\
& 850 &$^{b}$ & 26.7$\pm$1.4 &  & 51$^{c}$ & 2.3$^{c}$ &  \\
& 450 &$^{b}$ & 65$\pm$19 & & & &\\
& 350 &$^{dd}$ & $\sim 50$  & & &  &\\
SMM J02399 & 1270 &$^{ee}$ & 7.0$\pm$1.2 & & & & \\
& 1350 &$^{ff}$ & 5.7$\pm$1.0$^{}$ & \multirow{4}{*}{$\left. \phantom{\matrix{ \cr x \cr x \cr x  }} \right\}$}  & \multirow{4}{*}{11$^{ff}$} & \multirow{4}{*}{4.4$^{ff}$} & \multirow{4}{*}{6--8$^{ff}$} \\
& 850 &$^{ff}$ & 26$\pm$3 & & & & \\
& 750 &$^{ff}$ & 28$\pm$5 & & & & \\
& 450 &$^{ff}$ & 69$\pm$15 & & & & \\
SMM J04135 & 850 &$^{gg}$ & 25.0$\pm$2.8 & \multirow{2}{*}{$\left. \phantom{\matrix{ \cr x }} \right\}$} &\multirow{2}{*}{31$^{gg}$}  & \multirow{2}{*}{24$^{gg}$} & \multirow{2}{*}{18$^{gg}$} \\
& 450 &$^{gg}$ & 55$\pm$17 &  & & & \\
B3 J2330 & 1200 &$^{hh}$ & 4.8$\pm$1.2 & \multirow{2}{*}{$\left. \phantom{\matrix{ \cr x }} \right\}$} &\multirow{2}{*}{28$^{hh}$} &  \multirow{2}{*}{28$^{hh}$} &  \\
& 850 &$^{ii}$ & 14.1$\pm$1.7 & \multirow{2}{*}{$\left. \phantom{\matrix{ \cr x }} \right\}$} &\multirow{2}{*}{13$^{hh}$} &  \multirow{2}{*}{13$^{ii}$} &   \\
& 450 &$^{ii}$ & 49$\pm$18 & & & & \\
SMM J22174 &  850 &$^{l}$ & $6.3\pm1.3$ & & 12$^{c}$& & \\
MG0751 & 3000 &$^{jj}$ & 4.1$\pm$0.5 & & & & \\
& 1300 &$^{jj}$ & 6.7$\pm$1.3 & & & & \\
& 850 &$^{b}$ & 25.8$\pm$1.3 &  & 49$^{c}$ &   2.9$^{c}$ & \\
& 450 &$^{b}$ & 71$\pm$15$^{}$ & &  & & \\
SMM J09431 & 1300 &$^{m}$ & 2.3$\pm$0.4 & & &  &\\
& 850 &$^{kk}$ & $10.5\pm1.8$  &  & 20$^{m}$ &   17$^{m}$ & \\
SMM J13120 &  850 &$^{ll}$ & $6.2\pm1.2$ & & 12$^{c}$ & & \\
TN J0121 &  850 &$^{ii}$ & 7.5$\pm$1.0& & 7$^{ii}$ & 7$^{ii}$ & \\
6C1908 & 850 &$^{ii}$ & 10.8$\pm$1.2 & & 9.8$^{ii}$  & 9.8$^{ii}$ & \\
4C60.07 & 1250 &$^{mm}$ & 4.5$\pm$1.2 & & & & \\
& 850 &$^{ii}$ & 14.4$\pm$1.0 & & 13$^{ii}$ & 13$^{ii}$ &\\
& 850 &$^{nn}$ & 17.1$\pm$1.3 & & 32$^{c}$ & &\\
& 450 &$^{nn}$ & 69$\pm$23 & & & & \\
4C41.17 & 1245 &$^{oo}$ & 3.4$\pm$0.7 & &  \\
& 850 &$^{nn}$ & 12.1$\pm$0.9  &  &  &  & \\
& 450 &$^{nn}$ & 22.5$\pm$8.5  &  &  & & \\
& 350 &$^{j}$ & 37$\pm$9  &  &  20$^{j}$ & 20$^{j}$ & 4.6$^{j}$ \\
APM 08279 & 3200 &$^{pp}$ & 1.2$\pm$0.3 & & & & \\
& 1400 &$^{pp}$ & 17.0$\pm$0.5 & &  & &  \\
& 1300 &$^{b}$ & 24$\pm$2 & & & &  \\
&850 &$^{b}$ & 84$\pm$3 & &  &   & \\
& 450 &$^{b}$ & 285$\pm$11 & & &  & \\
& 350 &$^{y}$ & 392$\pm36$ &  & 200$^{y}$&  29$^{y}$ & 5.8$^{y}$\\
\end{tabular}
\end{table}

\begin{table} 
\footnotesize
\smallskip
\begin{tabular}{lr@{\tiny }llllcc}\hline\hline
\multicolumn{1}{c}{EMG} & \multicolumn{2}{c}{Band} & \multicolumn{1}{c}{Flux density} & \multicolumn{2}{c}{$L_{\rm FIR}(\rm app.)$} & \multicolumn{1}{c}{$L_{\rm FIR}(\rm int.)$}& $M_{\rm dust}$ \\
& \multicolumn{2}{c}{$(\mu \rm m)$} & \multicolumn{1}{c}{(mJy)} & \multicolumn{2}{c}{(10$^{12}$ L$_{\odot}$)} & \multicolumn{1}{c}{(10$^{12}$ L$_{\odot}$)} & ($10^{8}$ M$_{\odot}$) \\ \hline
PSS J2322 & 1200 &$^{qq}$ & 9.6$\pm$0.5 & & & & 9.0$^{qq}$\\
& 1350 &$^{rr}$ & 7.5$\pm$1.3 & \multirow{3}{*}{$\left. \phantom{\matrix{ \cr x \cr x }} \right\}$} &\multirow{3}{*}{23$^{rr}$}  &\multirow{3}{*}{9.3$^{rr}$} & \\
& 850 &$^{rr}$ & 24$\pm$2 & & & &\\
& 450 &$^{rr}$ & 79$\pm$19 & & &  &\\
& 350 &$^{y}$ & $66\pm9$& & 30$^{y}$&  12$^{y}$ & 9.6$^{y}$\\
BRI 1335 & 1350 &$^{cc}$ & 5.6$\pm$1.1 & & &  &\\
& 1250 &$^{cc}$ & 10.3$\pm$1.4 & & & & \\
& 350 &$^{j}$ & 52$\pm$8 & & 28$^{j}$ &  & 17$\mu^{-1}\ ^{j}$\\
BRI 0952 & 1350 &$^{cc}$ & 2.2$\pm$0.5 & & 9.6$^{cc}$ &  2.4$^{cc}$ & 0.7$^{cc}$\\
& 1250 &$^{cc}$ & 2.78$\pm$0.63 & & & & \\
& 850 &$^{b}$ & 13.4$\pm$2.3 & & 25$^{c}$ &  6.4$^{c}$ &\\
BR 1202 & 1350 &$^{cc}$ & 16$\pm$2 & & & &\\
& 350 &$^{j}$ & 106$\pm$7 & &71$^{j}$ & & 19$^{j}$ \\
SDSS J1148 & 1200 &$^{ss}$ & 5.0$\pm$0.6 & \multirow{3}{*}{$\left. \phantom{\matrix{ \cr x \cr x \cr x }} \right\}$} &\multirow{3}{*}{25$^{tt}$} & & 6.7$\mu^{-1}$ $^{ss}$\\
& 850 &$^{tt}$ & 7.8$\pm$0.7 & & & & 2.8$\mu^{-1}$ $^{tt}$\\
& 450 &$^{tt}$ & 24.7$\pm$7.4 & & &  &\\
& 350 &$^{y}$ & 23$\pm$3 & & 27$^{y}$ &  & 4.4$\mu^{-1}\ ^{y}$\\ \hline
\end{tabular}
\end{table}

\begin{table} 
\footnotesize
\smallskip
\begin{tabular}{lr@{\tiny }llllll}
\multicolumn{6}{l}{}\\ \hline
\multicolumn{6}{l}{\scriptsize{$^{a}$Smail et al. (2002);  $^{b}$Barvainis \& Ivison (2002); $^{c}$Using $\Lfir=1.9\times10^{12}S_{850}$, Neri et al. (2003);}}\\
\multicolumn{6}{l}{\scriptsize{$^{d}$Dey et al. (1999); $^{e}$Greve, Ivison \& Papadopoulos (1999); }}\\
\multicolumn{6}{l}{\scriptsize{$^{f}$D Downes \& PM Solomon, manuscript in preparation; $^{g}$Downes et al. (1992);}}\\
\multicolumn{6}{l}{\scriptsize{$^{h}$Rowan-Robinson et al. (1993); $^{i}$Clements et al. (1992); $^{j}$Benford et al. (1999);}}\\
\multicolumn{6}{l}{\scriptsize{$^{k}$Greve et al. (2004c); $^{l}$Chapman et al. (2005);$^{m}$Neri et al. (2003); $^{n}$Ivison et al. (2002); }}\\
\multicolumn{6}{l}{\scriptsize{$^{o}$Alloin, Barvainis \& Guilloteau (2000); $^{p}$Smail et al. (1999); $^{q}$Kneib et al. (2004a); }}\\
\multicolumn{6}{l}{\scriptsize{$^{r}$Kneib et al. (2004b); $^{s}$Sheth et al. (2004); $^{t}$Wei\ss\ et al. (2003); $^{u}$Downes \& Solomon (2003); }}\\
\multicolumn{6}{l}{\scriptsize{$^{v}$Mean for $\mu$=5--25, Downes \& Solomon (2003); $^{w}$Ivison et al. (2000); $^{x}$Omont et al. (2003);}}\\
\multicolumn{6}{l}{\scriptsize{$^{y}$A Beelen, P Cox, DJ Benford, CD Dowell, A Kovacs, et al., manuscript in preparation; }}\\
\multicolumn{6}{l}{\scriptsize{$^{z}$Priddey et al. (2003); $^{aa}$Baker (2001); $^{bb}$van der Werf et al. (2001); $^{cc}$Guilloteau et al. (1999);  }}\\
\multicolumn{6}{l}{\scriptsize{$^{dd}$J-W Wu, private communication; $^{ee}$Genzel et al. (2003); $^{ff}$Ivison et al. (1998); }}\\
\multicolumn{6}{l}{\scriptsize{$^{gg}$Knudsen, van der Werf \& Jaffe (2003); $^{hh}$De Breuck et al. (2003a); $^{ii}$Reuland et al. (2004);   }}\\
\multicolumn{6}{l}{\scriptsize{$^{jj}$Barvainis, Alloin \& Bremer (2002);  $^{kk}$Cowie, Barger \& Kneib (2002); $^{ll}$Chapman et al. (2003);  }}\\
\multicolumn{6}{l}{\scriptsize{$^{mm}$Papadopoulos et al. (2000); $^{nn}$Archibald et al. (2001); $^{oo}$De Breuck et al. (2004); }}\\
\multicolumn{6}{l}{\scriptsize{$^{pp}$Downes et al. (1999); $^{qq}$Omont et al. (2001); $^{rr}$Cox et al. (2002);  }}\\
\multicolumn{6}{l}{\scriptsize{$^{ss}$Bertoldi et al. (2003a); $^{tt}$Robson et al. (2004).}}\\
\end{tabular}
\end{table}



}
\end{document}